\documentclass[12pt,preprint]{aastex}

\usepackage{amsmath}
\usepackage{natbib}
\usepackage{graphicx}
\usepackage{color}
\usepackage[breaklinks,colorlinks,citecolor=blue]{hyperref}
\usepackage{epsfig}
\usepackage{epstopdf}

\usepackage{mathrsfs}

\begin{document}
\title{Kerr-Newman-AdS Black Hole Surrounded by Perfect Fluid Matter in Rastall Gravity}

\author{
  Zhaoyi Xu,\altaffilmark{1,2,3,4}
  Xian Hou,\altaffilmark{1,3,4}
  Xiaobo Gong,\altaffilmark{1,2,3,4}
  Jiancheng Wang \altaffilmark{1,2,3,4}
 }

\altaffiltext{1}{Yunnan Observatories, Chinese Academy of Sciences, 396 Yangfangwang, Guandu District, Kunming, 650216, P. R. China; {\tt zyxu88@ynao.ac.cn,xianhou.astro@gmail.com,xbgong@ynao.ac.cn,jcwang@ynao.ac.cn}}
\altaffiltext{2}{University of Chinese Academy of Sciences, Beijing, 100049, P. R. China}
\altaffiltext{3}{Key Laboratory for the Structure and Evolution of Celestial Objects, Chinese Academy of Sciences, 396 Yangfangwang, Guandu District, Kunming, 650216, P. R. China}
\altaffiltext{4}{Center for Astronomical Mega-Science, Chinese Academy of Sciences, 20A Datun Road, Chaoyang District, Beijing, 100012, P. R. China}

\shorttitle{Kerr-Newman-AdS Black Hole Surrounded by Perfect Fluid Matter in Rastall Gravity}
\shortauthors{Z Y. Xu et al.}

\begin{abstract}
The Rastall gravity is the modified Einstein general relativity, in which the energy-momentum conservation law is generalized to $T^{\mu\nu}_{~~;\mu}=\lambda R^{,\nu}$. In this work, we derive the Kerr-Newman-AdS (KN-AdS) black hole solutions surrounded by the perfect fluid matter in the Rastall gravity using the Newman-Janis method and Mathematica package. We then discuss the black hole properties surrounded by two kinds of specific perfect fluid matter, the dark energy ($\omega=-2/3$) and the perfect fluid dark matter ($\omega=-1/3$). Firstly, the Rastall parameter $\kappa\lambda$ could be constrained by the weak energy condition and strong energy condition. Secondly, by analyzing the number of roots in the horizon equation, we get the range of the perfect fluid matter intensity $\alpha$, which depends on the black hole mass $M$ and the Rastall parameter $\kappa\lambda$. Thirdly, we study the influence of the perfect fluid dark matter and dark energy on the ergosphere. We find that the perfect fluid dark matter has significant effects on the ergosphere size, while the dark energy has smaller effects. Finally, we find that the perfect fluid matter does not change the singularity of the black hole. Furthermore, we investigate the rotation velocity in the equatorial plane for the KN-AdS black hole with dark energy and perfect fluid dark matter. We propose that the rotation curve diversity in Low Surface Brightness galaxies could be explained in the framework of the Rastall gravity when both the perfect fluid dark matter halo and the baryon disk are taken into account.
\end{abstract}


\section{INTRODUCTION}
The dark matter and dark energy are two unsolved problems in cosmology and particle physics. Today many observations, including Type Ia supernovae, cosmic microwave background (CMB), baryon acoustic oscillations (BAO), weak lensing, rotation curve, larger scale structure etc., have revealed that the dark energy accounts for $73\%$, the dark matter for $23\%$ and ordinary baryonic matter only for $4\%$ of the total mass-energy of the universe \citep[e.g.,][]{2004ApJ...607..665R, 2011ApJS..192...18K}. Following these observations, many theoretical models have been proposed \citep[e.g.,][]{2003RvMP...75..559P, 2006IJMPD..15.1753C, 2017PhR...696....1W, 2011CoTPh..56..525L}. For example, the phenomenological dark energy models include quintessence, phantom, quintom and other models. The dark matter includes Cold Dark Matter (CDM) \citep[e.g.,][]{1991ApJ...378..496D, 1997ApJ...490..493N, 1996ApJ...462..563N}, Warm Dark Matter (WDM) and Scalar Field Dark Matter (SFDM) \citep[e.g.,][]{2005PhRvD..71f3534V, 2006PhLB..642..192A, 2000NewA....5..103G, 2000PhRvL..85.1158H, 2017PhRvD..95d3541H, 2016PhR...643....1M, 2000ApJ...534L.127P, 1990PhRvL..64.1084P, 2014NatPh..10..496S, 1994PhRvD..50.3650S, 1983PhRvD..28.1243T}. The perfect fluid dark matter (PFDM) model has also been proposed recently \citep[e.g.,][]{2003RMxF...49..203G, 2003gr.qc.....3031K, 2010PhLB..694...10R, 2011IJTP...50.2655R, 2017PhRvD..95f4015X}. For spiral galaxies, under the assumption of the spherical symmetric mass distribution and perfect fluid matter, one can obtain the solution of the Einstein equation for spherical symmetric gravitational field. When the equation of state for spiral galaxies is around $-1/3$ \citep{2003RMxF...49..203G}, the PFDM model can explain the asymptotically flat rotation curves.

The covariant conservation of the energy-momentum tensor plays an important role in Einstein's general relativity (GR). Taking this conservation as starting point, one can derive the conservation of some global defined physical quantities through Noether symmetry theorem. These conserved quantities take the form of integrals of the components of the energy-momentum tensor over appropriate space-like surfaces, which allow at least one of the Killing vectors of the background space-time as their normal. Thus, the total rest energy and rest mass of a physical system are conserved in the context of GR \citep{2017PhLB..771..365H}.
On the other hand, some modified Einstein GR have relaxed conservation condition of the energy-momentum tensor. One of the modified gravity theory was introduced by \cite{1972PhRvD...6.3357R, 1976CaJPh..54...66R}, in which the conservation of the energy-momentum tensor changes to be $T^{\mu\nu}_{~~;\mu}=\lambda R^{,\nu}$. When the space-time is flat, the conservation law becomes the usual formalism. This theory could be understood as Mach principle which represents the mass distribution depending on the energy and mass in the space-time \citep{2006gr.qc....10070M}, and has many interesting applications in cosmology and strong gravity space-time \citep[e.g.,][]{2012PhRvD..85h4008B, 2017arXiv171010429D, 2012IJMPS..18...67F, 2015AIPC.1647...50F, 2017PhLB..771..365H, 2016arXiv161003881H, 2017arXiv171004612L}. However,  the physical relevance of the Rastall gravity is under dispute. Recently, \cite{2017arXiv171111500V} pointed out that the Rastall gravity is equivalent to the Einstein GR, while \cite{2017arXiv171209307D} suggested that two gravitational theories are not equivalent, although both work admit that the parameter $\kappa\lambda$ always represents the rearrangement of the matter sector of the Einstein GR. In this context, it is still worth studying the Rastall gravity in both observational and theoretical aspects. 

The physical effects of the dark matter and dark energy on the black hole properties have been investigated recently and attracted increasing attention. The cosmological constant in Schwarzschild black hole space-time has been discussed in \cite{1918AnP...361..401K}, and later generalized to the rotational black hole by \cite{1973blho.conf...57C}. The dynamics of the quintessence dark energy in Schwarzschild space-time have been obtained by \cite{2003CQGra..20.1187K}, in which the equation of state $\omega$ for perfect fluid matter is constant. That work has been generalized to the Kerr black hole and Kerr-Newman-AdS (KN-AdS) black hole \citep{2015arXiv151201498T,2017PhRvD..95f4015X}. In the case of spherical symmetry, the Reissner-Nordstrom black hole space-time surrounded by perfect fluid matter in the Rastall gravity has been obtained by \cite{2017PhLB..771..365H}. In this work, we further generalize the Reissner-Nordstrom black hole space-time to the rotational black hole (KN-AdS black hole) space-time in the Rastall gravity and investigate accordingly the KN-AdS black hole properties.

The outline of the paper is as follows. In Section 2, we introduce the Reissner-Nordstrom black hole solution with perfect fluid matter in the Rastall gravity. In Section 3, we derive the KN-AdS black hole solution using Newman-Janis method and Mathematica package. In Section 4, we analyze the properties of the KN-AdS black hole surrounded by perfect fluid matter in the Rastall gravity, including the energy condition, horizon structure, ergosphere and singularity. In Section 5, we calculate the rotation velocity in the equatorial plane and discuss its applications. The summary is given in Section 6.

\section{REISSNER-NORDSTROM BLACK HOLE IN PERFECT FLUID MATTER}
In this section, we introduce the Reissner-Nordstrom black hole solution surrounded by perfect fluid matter in the Rastall gravity. The Rastall gravity is based on the Rastall hypothesis which generalizes the energy-momentum conservation law to the following formalism \citep{1972PhRvD...6.3357R,1976CaJPh..54...66R}
\begin{equation}
T^{\mu\nu}_{;\mu}=\lambda R^{,\nu},
\label{SY1}
\end{equation}
where $T_{\mu\nu}$ is the energy-momentum tensor, $\lambda$ is the Rastall parameter which represents the level of the energy-momentum conservation law in gravity theory and whether the Mach principle is satisfied. In this theory, the generalized Einstein field equation is given by
\begin{equation}
G_{\mu\nu}+\kappa\lambda g_{\mu\nu}R=\kappa T_{\mu\nu},
\label{SY2}
\end{equation}
where $\kappa=8\pi G_{N}$ is the gravitational constant of the Rastall gravity and $G_{N}$ is the gravitational constant of the Newton gravity. When $\lambda\rightarrow 0$, the field equation reduces to the Einstein field equation in GR.

In \cite{2017PhLB..771..365H}, the Reissner-Nordstrom black hole solution in perfect fluid matter is given by
\begin{equation}
ds^{2}=-f(r)dt^{2}+g^{-1}(r)dr^{2}+r^{2}(d\theta^{2}+sin^{2}\theta d\phi^{2}),
\label{SY3}
\end{equation}
where
\begin{equation}
f(r)=g(r)=1-\dfrac{2M}{r}+\dfrac{Q^{2}}{r^{2}}-\alpha r^{-\dfrac{1+3\omega-6\kappa\lambda(1+\omega)}{1-3\kappa\lambda(1+\omega)}}.
\label{SY4}
\end{equation}
where $\alpha$ is the perfect fluid matter intensity around the black hole, $M$ is the mass of the black hole, $Q$ is the charge of the black hole and $\omega$ describes the equation of state defined by $\omega=p/\rho$ with $p$ and $\rho$ the pressure and density of the perfect fluid matter, respectively. 

The electromagnetic field is encoded by the gauge potential $A$ which is
\begin{equation}
A=\dfrac{Q}{r}dt,
\label{KNADSq1}
\end{equation}
Although the real dark matter or dark energy could not satisfy the perfect fluid condition, the perfect fluid is a good approximation to describe the behaviour of the dark matter and dark energy. When $\lambda\rightarrow 0$, the black hole solution Eq. (\ref{SY3}) reduces to the Kiselev formalism \citep{2003CQGra..20.1187K}. For $-1<\omega<-1/3$, the solution represents the black hole surrounded by the dark energy \citep{2003CQGra..20.1187K,2015arXiv151201498T,2017PhRvD..95f4015X}. For $\omega=-1/3$, the solution represents the black hole surrounded by the perfect fluid dark matter \citep{2003RMxF...49..203G, 2003gr.qc.....3031K, 2010PhLB..694...10R, 2011IJTP...50.2655R, 2017PhRvD..95f4015X}.

\section{KN-ADS BLACK HOLE IN PERFECT FLUID MATTER}
\subsection{Newman-Janis Method and Kerr-Newman Solution in Perfect Fluid Matter}
Using the Newman-Janis method \citep{2014PhRvD..90f4041A, 2010CQGra..27p5008C, 2015GReGr..47...19E, 1965JMP.....6..915N}, we generalize the Reissner-Nordstrom black hole to the Kerr-Newman black hole in perfect fluid matter in the Rastall gravity. First, we perform transformation for the spherically symmetric black hole space-time metric (Eq. \ref{SY3}) from Boyer-Lindquist (BL) coordinates $(t,r,\theta,\phi)$ to Eddington-Finkelstein (EF) coordinates $(u,r,\theta,\phi)$ through
\begin{equation}
du=dt-\dfrac{dr}{1-\dfrac{2M}{r}+\dfrac{Q^{2}}{r^{2}}-\alpha r^{-\dfrac{1+3\omega-6\kappa\lambda(1+\omega)}{1-3\kappa\lambda(1+\omega)}}},
\label{NJ1}
\end{equation}
and the space-time metric (Eq. \ref{SY3}) becomes
\begin{equation}
ds^{2}=-(1-\dfrac{2M}{r}+\dfrac{Q^{2}}{r^{2}}-\alpha r^{-\dfrac{1+3\omega-6\kappa\lambda(1+\omega)}{1-3\kappa\lambda(1+\omega)}})du^{2}-2dudr+r^{2}d\Omega^{2}.
\label{NJ2}
\end{equation}
For the electromagnetic field part, the gauge potential $A$ in the $(u,r)$ plane becomes 
\begin{equation}
A=\dfrac{Q}{r}(du+\dfrac{dr}{1-\dfrac{2M}{r}+\dfrac{Q^{2}}{r^{2}}-\alpha r^{-\dfrac{1+3\omega-6\kappa\lambda(1+\omega)}{1-3\kappa\lambda(1+\omega)}}}).
\label{KNADSq2}
\end{equation}
The non-zero components of the inverse space-time metric (Eq. \ref{NJ2}) are given by
\begin{equation}
g^{rr}=1-\dfrac{2M}{r}+\dfrac{Q^{2}}{r^{2}}-\alpha r^{-\dfrac{1+3\omega-6\kappa\lambda(1+\omega)}{1-3\kappa\lambda(1+\omega)}},~~~    g^{\theta\theta}=\dfrac{1}{r^{2}}, $$$$  g^{\phi\phi}=\dfrac{1}{r^{2}sin^{2}\theta}, ~~~ g^{ur}=g^{ru}=-1.
\label{NJ3}
\end{equation}
In the null (EF) frame, the metric matrix can be written as
\begin{equation}
g^{\mu\nu}=-l^{\mu}n^{\nu}-l^{\nu}n^{\mu}+m^{\mu}\overline{m}^{\nu}+m^{\nu}\overline{m}^{\mu},
\label{NJ4}
\end{equation}
where the corresponding components are
\begin{equation}
l^{\mu}=\delta^{\mu}_{r},$$$$
n^{\mu}=\delta^{\mu}_{0}-\dfrac{1}{2}(1-\dfrac{2M}{r}+\dfrac{Q^{2}}{r^{2}}-\alpha r^{-\dfrac{1+3\omega-6\kappa\lambda(1+\omega)}{1-3\kappa\lambda(1+\omega)}})\delta^{\mu}_{r},$$$$
m^{\mu}=\dfrac{1}{\sqrt{2}r}\delta^{\mu}_{\theta}+\dfrac{i}{\sqrt{2}r sin\theta}\delta^{\mu}_{\phi},$$$$
\overline{m}^{\mu}=\dfrac{1}{\sqrt{2}r}\delta^{\mu}_{\theta}-\dfrac{i}{\sqrt{2}r sin\theta}\delta^{\mu}_{\phi}.
\label{NJ5}
\end{equation}
For any point in the black hole space-time, the null vectors of the null tetrad satisfy the relations of $l_{\mu}l^{\mu}=n_{\mu}n^{\mu}=m_{\mu}m^{\mu}=l_{\mu}m^{\mu}=n_{\mu}m^{\mu}=0$ and $l_{\mu}n^{\mu}=-m_{\mu}\overline{m}^{\mu}=1$. In the plane of ($u, r$), the coordinate transformations are
\begin{equation}
u\longrightarrow u-iacos\theta, $$$$
r\longrightarrow r-iacos\theta.
\label{NJ6}
\end{equation}
We then perform the transformations of $f(r)\rightarrow F(r,a,\theta)$, $g(r)\rightarrow G(r,a,\theta)$, and $\Sigma^{2}=r^{2}+a^{2}cos^{2}\theta$. In the ($u, r$) space, the null vectors become
\begin{equation}
l^{\mu}=\delta^{\mu}_{r},~~~~n^{\mu}=\sqrt{\dfrac{G}{F}}\delta^{\mu}_{0}-\dfrac{1}{2}F\delta^{\mu}_{r},$$$$
m^{\mu}=\dfrac{1}{\sqrt{2}\Sigma}(\delta^{\mu}_{\theta}+ia sin\theta(\delta^{\mu}_{0}-\delta^{\mu}_{r})+\dfrac{i}{sin\theta}\delta^{\mu}_{\phi}),$$$$
\overline{m}^{\mu}=\dfrac{1}{\sqrt{2}\Sigma}(\delta^{\mu}_{\theta}-ia sin\theta(\delta^{\mu}_{0}-\delta^{\mu}_{r})-\dfrac{i}{sin\theta}\delta^{\mu}_{\phi}).
\label{NJ7}
\end{equation}
From the definition of the null tetrad, the metric tensor $g^{\mu\nu}$ in the EF coordinates are given by
\begin{equation}
g^{uu}=\dfrac{a^{2}sin^{2}\theta}{\Sigma^{2}},~~~~~~g^{rr}=G+\dfrac{a^{2}sin^{2}\theta}{\Sigma^{2}},$$$$
g^{\theta\theta}=\dfrac{1}{\Sigma^{2}},~~~~~~g^{\phi\phi}=\dfrac{1}{\Sigma^{2}sin^{2}\theta},$$$$
g^{ur}=g^{ru}=-\sqrt{\dfrac{G}{F}}-\dfrac{a^{2}sin^{2}\theta}{\Sigma^{2}},$$$$
g^{u\phi}=g^{\phi u}=\dfrac{a}{\Sigma^{2}},~~~~~~g^{r\phi}=g^{\phi r}=-\dfrac{a}{\Sigma^{2}}.
\label{NJ8}
\end{equation}
Through calculation, the covariant metric tensor in the EF coordinates are
\begin{equation}
g_{uu}=-F,~~~~~~g_{\theta\theta}=\Sigma^{2},~~~~~~g_{ur}=g_{ru}=-\sqrt{\dfrac{G}{F}},$$$$
g_{\phi\phi}=sin^{2}\theta(\Sigma^{2}+a^{2}(2\sqrt{\dfrac{F}{G}}-F)sin^{2}\theta),$$$$
g_{u\phi}=g_{\phi u}=a(F-\sqrt{\dfrac{F}{G}})sin^{2}\theta,~~~~~~g_{r\phi}=g_{\phi r}=a sin^{2}\theta\sqrt{\dfrac{F}{G}}.
\label{NJ9}
\end{equation}
Using Eq.12 we can obtain the electromagnetic field gauge potential $A$ in the EF coordinates as 
\begin{equation}
A=\dfrac{Qr}{\Sigma^{2}}(du-a sin^{2}\theta d\phi).
\label{KNADSq3}
\end{equation}
Finally we perform the coordinate transformation from the EF to BL coordinates as
\begin{equation}
du=dt+\lambda(r)dr,~~~~d\phi=d\phi+h(r)dr,
\label{NJ10}
\end{equation}
where
\begin{equation}
\lambda(r)=-\dfrac{r^{2}+a^{2}}{r^{2}g(r)+a^{2}},~~~h(r)=-\dfrac{a}{r^{2}g(r)+a^{2}},~~~
F(r,\theta)=G(r,\theta)=\dfrac{r^{2}g(r)+a^{2}cos^{2}\theta}{\Sigma^{2}}.
\label{NJ11}
\end{equation}
In the BL coordinates, the Kerr-Newman black hole solution in perfect fluid matter is therefore given by
\begin{equation}
ds^{2}=-(1-\dfrac{2Mr-Q^{2}+\alpha r^{\dfrac{1-3\omega}{1-3\kappa\lambda(1+\omega)}}}{\Sigma^{2}})dt^{2}-\dfrac{2a sin^{2}\theta(2Mr-Q^{2}+\alpha r^{\dfrac{1-3\omega}{1-3\kappa\lambda(1+\omega)}})}{\Sigma^{2}}d\phi dt$$$$
+\Sigma^{2}d\theta^{2}+\dfrac{\Sigma^{2}}{\Delta_{r}}dr^{2}+sin^{2}\theta (r^{2}+a^{2}+a^{2}sin^{2}\theta\dfrac{2Mr-Q^{2}+\alpha r^{\dfrac{1-3\omega}{1-3\kappa\lambda(1+\omega)}}}{\Sigma^{2}})d\phi^{2},
\label{NJ12}
\end{equation}
where
\begin{equation}
\Delta_{r}=r^{2}-2Mr+a^{2}+Q^{2}-\alpha r^{\dfrac{1-3\omega}{1-3\kappa\lambda(1+\omega)}},
\label{NJ13}
\end{equation}
and the gauge potential $A$ is accordingly  
\begin{equation}
A=\dfrac{Qr}{\Sigma^{2}}(dt-a sin^{2}\theta d\phi).
\label{KNADSq4}
\end{equation}
From the Newman-Janis method, we know that this Kerr-Newman black hole space-time metric in the Rastall gravity (Eq. 19) satisfies the Einstein-Maxwell field equation. If there is no perfect fluid matter around the black hole ($\alpha=0$), this solution will reduce to the usual Kerr-Newman black hole without dark matter. If $\kappa\lambda=0$, it will reduce to the Kerr-Newman black hole in perfect fluid matter in the Einstein GR \citep{2017PhRvD..95f4015X}.

\subsection{KN-AdS Solution in Perfect Fluid Matter}
Now we extend the Kerr-Newman black hole to the KN-AdS black hole surrounded by the perfect fluid matter. Since the Newman-Janis method does not include the cosmological constant $\Lambda$, we employ other method to obtain the solution with the presence of the cosmological constant, as presented in \cite{2017PhRvD..95f4015X}. First, we rewrite the above Kerr-Newman black hole metric (Eq. 19) as
\begin{equation}
ds^{2}=\dfrac{\Sigma^{2}}{\Delta_{r}}dr^{2}+\Sigma^{2}d\theta^{2}+\dfrac{sin^{2}\theta}{\Sigma^{2}}(adt-(r^{2}+a^{2})d\phi)^{2}-\dfrac{\Delta_{r}}{\Sigma^{2}}(dt-a sin^{2}d\phi)^{2}.
\label{AdS1}
\end{equation}

We then guess the KN-AdS black hole solution as
\begin{equation}
ds^{2}=\dfrac{\Sigma^{2}}{\Delta_{r}}dr^{2}+\dfrac{\Sigma^{2}}{\Delta_{\theta}}d\theta^{2}+\dfrac{\Delta_{\theta}sin^{2}\theta}{\Sigma^{2}}(a\dfrac{dt}{\Xi}-(r^{2}+a^{2})\dfrac{d\phi}{\Xi})^{2}-\dfrac{\Delta_{r}}{\Sigma^{2}}(\dfrac{dt}{\Xi}-a sin^{2}\dfrac{d\phi}{\Xi})^{2},
\label{AdS5}
\end{equation}
where
\begin{equation}
\Delta_{r}=r^{2}-2Mr+a^{2}+Q^{2}-\alpha r^{\dfrac{1-3\omega}{1-3\kappa\lambda(1+\omega)}}-\dfrac{\Lambda}{3}r^{2}(r^{2}+a^{2}),$$$$
\Delta_{\theta}=1+\dfrac{\Lambda}{3}a^{2}cos^{2}\theta,~~~~~~~~\Xi=1+\dfrac{\Lambda}{3}a^{2}.
\label{AdS6}
\end{equation}

The solution should satisfy the Einstein-Maxwell field equations which are
\begin{equation}
G_{\mu\nu}=R_{\mu\nu}-\dfrac{1}{2}R g_{\mu\nu}+\kappa\lambda g_{\mu\nu}R+\Lambda g_{\mu\nu}=\kappa T_{\mu\nu},$$$$
F^{\mu\nu}_{;\nu}=0; ~~~ F^{\mu\nu;\alpha}+F^{\nu\alpha;\mu}+F^{\alpha\mu;\nu}=0.
\label{AdS11}
\end{equation}

Through calculations using the Mathematica package ( inserting Eq. (\ref{AdS5}) into $G_{\mu\nu}$ (Eq. 25)), we obtain the Einstein tensors in the Rastall gravity as
\begin{equation}
G_{tt}=\dfrac{1}{\Sigma^{6}}[\dfrac{Q^{2}}{2r^{2}}+\dfrac{\alpha (3\kappa\lambda(1+\omega)-3\omega)}{2(1-3\kappa\lambda(1+\omega))}r^{\dfrac{6\kappa\lambda(1+\omega)-3\omega-1}{1-3\kappa\lambda (1+\omega)}}][r^{4}-2r^{3}M+r^{2}Q^{2}-\alpha r^{\dfrac{6\kappa\lambda(1+\omega)-3\omega+3}{1-3\kappa\lambda(1+\omega)}}$$$$
+a^{2}r^{2}-a^{4}sin^{2}\theta cos^{2}\theta]-\dfrac{r a^{2}sin^{2}\theta}{\Sigma^{4}}[-\dfrac{Q^{2}}{r^{3}}+\dfrac{\alpha [3\kappa\lambda(1+\omega)-3\omega][6\kappa\lambda(1+\omega)-3\omega-1]}{2(1-3\kappa\lambda(1+\omega))^{2}}r^{\dfrac{9\kappa\lambda(1+\omega)-3\omega-2}{1-3\kappa\lambda(1+\omega)}}]$$$$
=\dfrac{2[r^{4}-2r^{3}W+a^{2}r^{2}-a^{4}sin^{2}\theta cos^{2}\theta]W^{'}}{\Sigma^{6}}-\dfrac{ra^{2}sin^{2}\theta W^{''}}{\Sigma^{4}},$$$$
G_{rr}=\dfrac{1}{\Sigma^{2}\Delta_{r}}[-Q^{2}-\alpha\dfrac{3\kappa\lambda(1+\omega)-3\omega}{1-3\kappa\lambda(1+\omega)}r^{\dfrac{1-3\omega}{1-3\kappa\lambda(1+\omega)}}]=-\dfrac{2r^{2}W^{'}}{\Sigma^{2}\Delta_{r}},$$$$
G_{t\phi}=\dfrac{2a sin^{2}\theta[(r^{2}+a^{2})(a^{2}cos^{2}\theta-r^{2})]}{\Sigma^{6}}[\dfrac{Q^{2}}{2r^{2}}+\dfrac{\alpha(3\kappa\lambda(1+\omega)-3\omega)}{2(1-3\kappa\lambda(1+\omega))}r^{\dfrac{6\kappa\lambda(1+\omega)-3\omega-1}{1-3\kappa\lambda(1+\omega)}}]$$$$
-\dfrac{ra^{2}sin^{2}\theta(r^{2}+a^{2})}{\Sigma^{4}}[-\dfrac{Q^{2}}{r^{3}}+\dfrac{\alpha [3\kappa\lambda(1+\omega)-3\omega][6\kappa\lambda(1+\omega)-3\omega-1]}{2(1-3\kappa\lambda(1+\omega))^{2}}r^{\dfrac{9\kappa\lambda(1+\omega)-3\omega-2}{1-3\kappa\lambda(1+\omega)}}]$$$$
=\dfrac{2a sin^{2}\theta[(r^{2}+a^{2})(a^{2}cos^{2}\theta-r^{2})]W^{'}}{\Sigma^{6}}-\dfrac{ra^{2}sin^{2}\theta(r^{2}+a^{2})W^{''}}{\Sigma^{4}},$$$$
G_{\theta\theta}=-\dfrac{2a^{2}cos^{2}\theta}{\Sigma^{2}}[\dfrac{Q^{2}}{2r^{2}}+\dfrac{\alpha(3\kappa\lambda(1+\omega)-3\omega)}{2(1-3\kappa\lambda(1+\omega))}r^{\dfrac{6\kappa\lambda(1+\omega)-3\omega-1}{1-3\kappa\lambda(1+\omega)}}]-r[-\dfrac{Q^{2}}{r^{3}}+$$$$
\dfrac{\alpha [3\kappa\lambda(1+\omega)-3\omega][6\kappa\lambda(1+\omega)-3\omega-1]}{2(1-3\kappa\lambda(1+\omega))^{2}}r^{\dfrac{9\kappa\lambda(1+\omega)-3\omega-2}{1-3\kappa\lambda(1+\omega)}}]
=-\dfrac{2a^{2}cos^{2}\theta W^{'}}{\Sigma^{2}}-rW^{''},$$$$
G_{\phi\phi}=-\dfrac{a^{2}sin^{2}\theta}{\Sigma^{6}}[(r^{2}+a^{2})(a^{2}+(2r^{2}+a^{2})cos2\theta)+2r^{3}sin^{2}\theta (M-\dfrac{Q^{2}}{2r}+\dfrac{\alpha}{2}r^{\dfrac{3\kappa\lambda(1+\omega)-3\omega}{1-3\kappa\lambda(1+\omega)}}))][\dfrac{Q^{2}}{2r^{2}}+$$$$
\dfrac{\alpha(3\kappa\lambda(1+\omega)-3\omega)}{2(1-3\kappa\lambda(1+\omega))}r^{\dfrac{6\kappa\lambda(1+\omega)-3\omega-1}{1-3\kappa\lambda(1+\omega)}}]-\dfrac{rsin^{2}\theta(r^{2}+a^{2})^{2}}{\Sigma^{4}}[-\dfrac{Q^{2}}{r^{3}}+$$$$
\dfrac{\alpha [3\kappa\lambda(1+\omega)-3\omega][6\kappa\lambda(1+\omega)-3\omega-1]}{2(1-3\kappa\lambda(1+\omega))^{2}}r^{\dfrac{9\kappa\lambda(1+\omega)-3\omega-2}{1-3\kappa\lambda(1+\omega)}}]
$$$$
=-\dfrac{a^{2}sin^{2}\theta[(r^{2}+a^{2})(a^{2}+(2r^{2}+a^{2})cos2\theta)+2r^{3}sin^{2}\theta W)]W^{'}}{\Sigma^{6}}-\dfrac{rsin^{2}\theta(r^{2}+a^{2})^{2}W^{''}}{\Sigma^{4}},
\label{AdSZ7}
\end{equation}
where $W(r)=M-\dfrac{Q^{2}}{2r}+\dfrac{\alpha}{2}r^{\dfrac{3\kappa\lambda(1+\omega)-3\omega}{1-3\kappa\lambda(1+\omega)}}$. 

Generically the Kerr black hole in GR is a vacuum solution and hence the energy-momentum tensor is zero. A non-zero cosmological constant $\Lambda$ in GR incorporates a static geometry which is known to only shift its vacuum energy by a constant, while has no effect on the energy-momentum tensor. Analogically, in the Rastall gravity, the non-zero $\Lambda$ should not appear in the explicit expression of the Einstein tensor (or energy-momentum tensor) components either, in the case of the KN-AdS black hole space-time surrounded by the perfect fluid matter. This is exactly what we obtained (Eq. 26). Therefore, our guessed solution (Eq. \ref{AdS5}) satisfies the Einstein-Maxwell field equation (Eq. 25) with the perfect fluid matter in the Rastall gravity. Similarly, the gauge potential $A$ in the KN-AdS black hole does not involve $\Lambda$ and has same expression as in the case of Kerr-Newman black hole (Eq. 21), therefore it also satisfies the Einstein-Maxwell equation (Eq. 25). 


In order to compute the energy-momentum tensor, we give the following tetrad
\begin{equation}
e^{\mu}_{t}=\dfrac{1}{\sqrt{\Xi^{2}\Sigma^{2}\Delta_{r}}}(r^{2}+a^{2},0,0,a),$$$$
e^{\mu}_{r}=\dfrac{\sqrt{\Delta_{r}}}{\sqrt{\Sigma^{2}}}(0,1,0,0),$$$$
e^{\mu}_{\theta}=\dfrac{\sqrt{\Delta_{\theta}}}{\sqrt{\Sigma^{2}}}(0,0,1,0),$$$$
e^{\mu}_{\phi}=-\dfrac{1}{\sqrt{\Xi^{2}\Sigma^{2}sin^{2}\theta}}(r^{2}+a^{2},0,0,a).
\label{EM1}
\end{equation}
The non-zero components of the energy-momentum tensor are given by
\begin{equation}
E=-\dfrac{1}{\kappa}e^{\mu}_{t}e^{\nu}_{t}G_{\mu\nu}=-\dfrac{1}{\kappa}g^{tt}(R_{tt}-\dfrac{1}{2}Rg_{tt}+\kappa\lambda g_{tt}R+\Lambda g_{tt}),$$$$
P_{r}=\dfrac{1}{\kappa}e^{\mu}_{r}e^{\nu}_{r}G_{\mu\nu}=\dfrac{1}{\kappa}g^{rr}(R_{rr}-\dfrac{1}{2}Rg_{rr} +\kappa\lambda g_{rr}R+\Lambda g_{rr}),$$$$
P_{\theta}=\dfrac{1}{\kappa}e^{\mu}_{\theta}e^{\nu}_{\theta}G_{\mu\nu}=\dfrac{1}{\kappa}g^{\theta\theta}(R_{\theta\theta}-\dfrac{1}{2}Rg_{\theta\theta} +\kappa\lambda g_{\theta\theta}R+\Lambda g_{\theta\theta}),$$$$
P_{\phi}=-\dfrac{1}{\kappa}e^{\mu}_{\phi}e^{\nu}_{\phi}G_{\mu\nu}=-\dfrac{1}{\kappa}g^{\phi\phi}(R_{\phi\phi}-\dfrac{1}{2}Rg_{\phi\phi} +\kappa\lambda g_{\phi\phi}R+\Lambda g_{\phi\phi}).
\label{EM2}
\end{equation}
We then find that
\begin{equation}
\rho=E=-P_{r}=\dfrac{r^{2}}{\kappa(r^{2}+a^{2}cos^{2}\theta)^{2}}
\dfrac{\alpha(3\kappa\lambda(1+\omega)-3\omega)(1-4\kappa\lambda)}{(1-3\kappa\lambda(1+\omega))^{2}}r^{\dfrac{6\kappa\lambda(1+\omega)-3\omega-1}{1-3\kappa\lambda(1+\omega)}},$$$$
P_{\theta}=P_{\phi}=-P_{r}-$$$$
\dfrac{\alpha [3(\kappa\lambda-\omega)+9\omega^{2}(1-5\kappa\lambda)+6\omega \kappa\lambda+6\kappa\lambda^{2}(1+\omega)(6\omega-2-6\omega \kappa\lambda)]}{2\kappa(r^{2}+a^{2}cos^{2}\theta)(1-3\kappa\lambda(1+\omega))^{2}}r^{\dfrac{6\kappa\lambda(1+\omega)-3\omega-1}{1-3\kappa\lambda(1+\omega)}}.
\label{EM3}
\end{equation}

It is clear that the distribution of the perfect fluid matter around the black hole (Eq. 29) is determined by the parameters of $a$, $\Lambda$, $Q$, $\alpha$, $\omega$ and $\kappa\lambda$. The space-time metric (Eq. \ref{AdS5}) describes the perfect fluid matter around the KN-AdS black hole in the Rastall gravity. The equation of state $\omega$ determines the material form. As described in Section 2, if $-1<\omega<-1/3$, the perfect fluid matter represents the dark energy. If $\omega=-1/3$, the perfect fluid matter stands for the perfect fluid dark matter. Using this space-time metric, we can study the interaction between the perfect fluid dark matter (or dark energy) and black hole. Such topic is very interesting in regard of the dark matter (dark energy)-black hole systems. From Eq. (29), we can see that the Rastall parameter $\kappa\lambda$ plays an important role in the distribution of perfect fluid matter around the black hole. Specifically, the Rastall gravity makes the perfect fluid matter to be rearranged around the black hole. If the Rastall gravity is equivalent to the Einstein GR, $\kappa\lambda$ only describes the re-arrangement of the perfect fluid matter. If the Rastall gravity is not equivalent to the Einstein GR, the meaning of $\kappa\lambda$ is the same, but the physical origin of $\kappa\lambda$ is related to the Mach principle.

\section{KN-ADS BLACK HOLE PROPERTIES IN PERFECT FLUID MATTER}

\subsection{Energy condition}
In this part, we study the energy condition of the KN-AdS black hole in perfect fluid matter in the Rastall gravity. First, we consider the diagonal energy-momentum tensor $T_{\mu\nu}$ in the standard locally non-rotating frame \citep[LNRF,][]{1972ApJ...178..347B}. The weak energy condition (WEC) can be written as $T_{\mu\nu}u^{\mu}u^{\nu}\geq 0$, where $u^{\nu}$ is the time-like vector. The physical meaning of the WEC is that the measured total energy density of all matter fields for any observer traversing a time-like curve is never negative. In the LNRF, the WEC corresponds to $\rho\geq 0$ and $\rho+P_{i}\geq 0$, where $i=r,\theta,\phi$. From Eq. 29 the WEC requires that 
\begin{equation}
\alpha(3\kappa\lambda(1+\omega)-3\omega)(1-4\kappa\lambda)\geq 0,
\label{EC1}
\end{equation}
which is independent of the black hole spin $a$. Thus, the WEC for the KN-AdS black hole in perfect fluid matter in the Rastall gravity is the same as that in the case of the spherically symmetric black hole \citep{2017PhLB..771..365H}. Since $\alpha$ is always positive, the condition $(3\kappa\lambda(1+\omega)-3\omega)(1-4\kappa\lambda)\geq 0$ will always be satisfied. If considering dark energy ($\omega=-2/3$), then this condition will result in $-2\leq\kappa\lambda\leq 1/4$. If considering dark matter ($\omega=-1/3$), then $-1/2\leq\kappa\lambda\leq 1/4$.

The strong energy condition (SEC) requires that the Raychaudhuri equation is $(T_{\mu\nu}-\dfrac{1}{2}Tg_{\mu\nu})u^{\mu}u^{\nu}\geq 0$. In the LNRF, the Raychaudhuri equation becomes $Y=\rho+P_{r}+P_{\theta}+P_{\phi}$. From (Eq. 29) the SEC implies that
\begin{equation}
Y=\rho+P_{r}+P_{\theta}+P_{\phi}=\dfrac{\alpha}{\kappa\Sigma^{4}(1-3\kappa\lambda(1+\omega))^{2}}[2r^{2}(3\kappa\lambda(1+\omega)-3\omega)(1-4\kappa\lambda)-\Sigma^{2}(3(\kappa\lambda-\omega)$$$$
+9\omega^{2}(1-5\kappa\lambda)+6\omega \kappa\lambda+6\kappa\lambda^{2}(1+\omega)(6\omega-2-6\omega \kappa\lambda))]\geq 0,
\label{EC2}
\end{equation}
This equation describes the condition that $\alpha$ and $\kappa\lambda$ should satisfy when SEC is satisfied. From Eq . 31 we find that $\rho+P_{r}+P_{\theta}+P_{\phi}=0$ has one zero point, then $Y=\rho+P_{r}+P_{\theta}+P_{\phi}$ is in the range $[r_{1},\infty)$ in the KN-AdS space-time with perfect fluid matter in the Rastall gravity (where $r_{1}$ is the root of Eq . 31), but at the same time, $\kappa\lambda$ must satisfy the condition $(3\kappa\lambda(1+\omega)-3\omega)(1-4\kappa\lambda)\geq 0$ which is similar to the WEC condition (Eq . 30). Therefore, the range of $\kappa\lambda$ under the SEC is consistent with the case of WEC.

\subsection{Horizon structure}
Black hole properties are determined by the horizon structure of the black hole. For stationary axisymmetric black hole, the horizon definition is
\begin{equation}
\Delta_{r}=r^{2}-2Mr+a^{2}+Q^{2}-\alpha r^{\dfrac{1-3\omega}{1-3\kappa\lambda(1+\omega)}}-\dfrac{\Lambda}{3}r^{2}(r^{2}+a^{2})=0,
\label{EH1}
\end{equation}
where the property of horizon depends on $\alpha, a, Q, \Lambda, \kappa\lambda$ and $\omega$. Especially the parameters of $\kappa\lambda$ and $\omega$ change the horizon significantly. The number of horizon is determined by $\omega$. As examples, we discuss the cases of the dark energy ($\omega=-2/3$) and dark matter ($\omega=-1/3$). We can set the cosmological constant $\Lambda=0$ given that it is small.

Case I: for the dark energy, the horizon equation reduces to
\begin{equation}
r^{2}-2Mr+a^{2}+Q^{2}-\alpha r^{\dfrac{3}{1-\kappa\lambda}}=0.
\label{EH2}
\end{equation}
Three horizons exist, including inner horizon ($r_{-}$), event horizon ($r_{+}$) and cosmological horizons ($r_{q}$), where $r_{q}$ is determined by the dark energy. Then the above equation should have two extreme value points. Through calculation, we find that $\alpha$ satisfies the condition of
\begin{equation}
0<\alpha\leq \dfrac{(1-\kappa\lambda)^{2}}{3(2+\kappa\lambda)}(2M)^{\dfrac{-\kappa\lambda}{1-\kappa\lambda}}.
\label{EH3}
\end{equation}
When $M=1$ and $\kappa\lambda=0$, the above condition reduces to the Kiselev situation. In other cases, the maximum value of $\alpha$ decreases with the increasing $\kappa\lambda$ when $0<\kappa\lambda$ and $\kappa\lambda\neq 1$, and increases with the increasing of $\mid\kappa\lambda\mid$ when $-2<\kappa\lambda<0$.

Case II: for the perfect fluid dark matter, the horizon equation reduces to
\begin{equation}
r^{2}-2Mr+a^{2}+Q^{2}-\alpha r^{\dfrac{2}{1-2\kappa\lambda}}=0.
\label{EH10}
\end{equation}
Dark matter does not produce new horizon and only two normal horizons exist, i.e., inner horizon $r_{-}$ and event horizon $r_{+}$. Then the horizon equation should have one extreme value point. Through calculation, we find that $\alpha$ satisfies the condition of
\begin{equation}
0<\alpha\leq (1-\kappa\lambda)(2M)^{-\dfrac{4\kappa\lambda}{1-2\kappa\lambda}}.
\label{EH11}
\end{equation}
When $M=1$ and $\kappa\lambda=0$, the above condition reduces to the Kiselev situation. Similarly, the maximum value of $\alpha$ decreases with the increasing $\kappa\lambda$ when $0<\kappa\lambda<1$ and $\kappa\lambda\neq 1/2$, and increases with the increasing $\mid\kappa\lambda\mid$ when $\kappa\lambda<0$.

These results reveal that $\kappa\lambda$ would constrain $\alpha$ in the Rastall gravity. As mentioned in Section 1 and 3, the physical origin could have two possibilities. If the Rastall gravity is equivalent to the Einstein GR, $\kappa\lambda$ only describes the re-arrangement of the perfect fluid matter. If the Rastall gravity is not equivalent to the Einstein GR, the meaning of $\kappa\lambda$ is the same, but its physical origin is related to the Mach principle. Furthermore, the Mach principle would determine the value of $\alpha$.

\begin{figure}[htbp]
  \centering
  \includegraphics[scale=0.36]{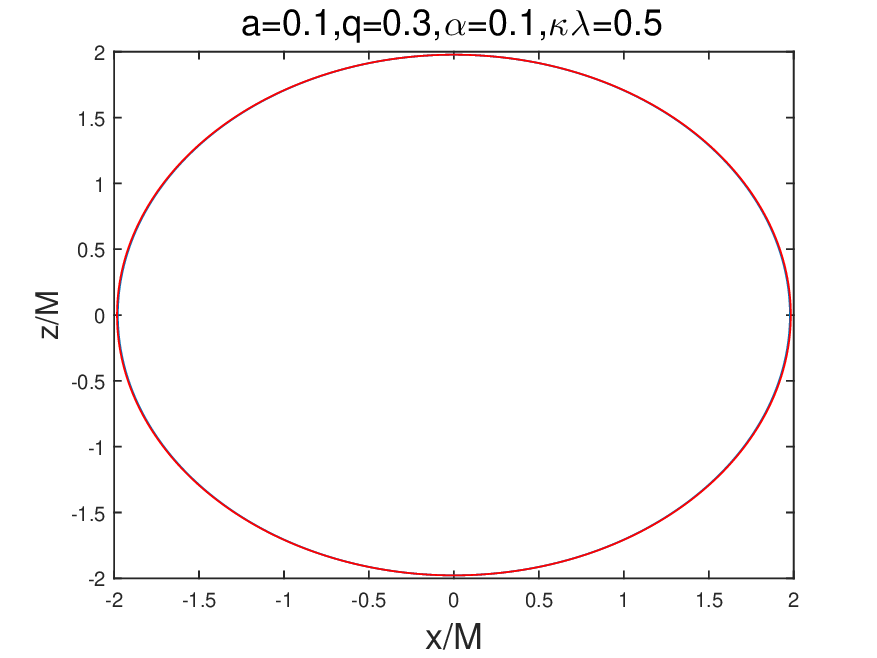}
  \includegraphics[scale=0.36]{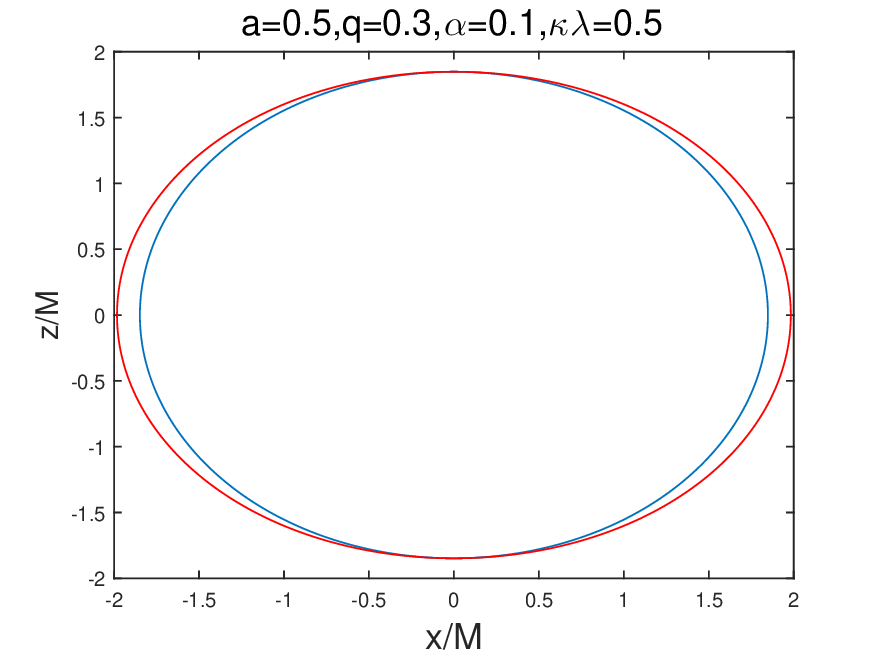}
  \includegraphics[scale=0.36]{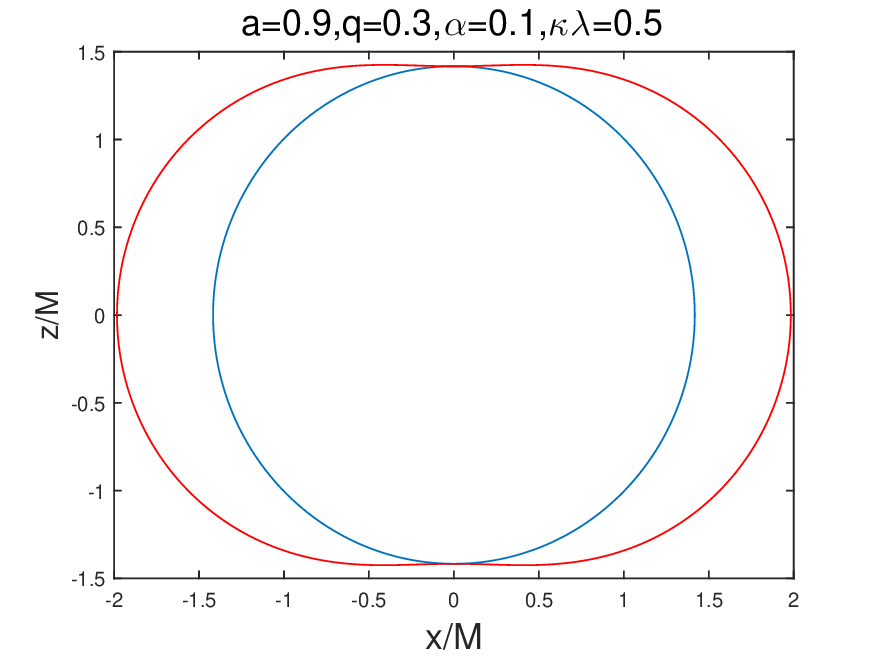}
  \includegraphics[scale=0.36]{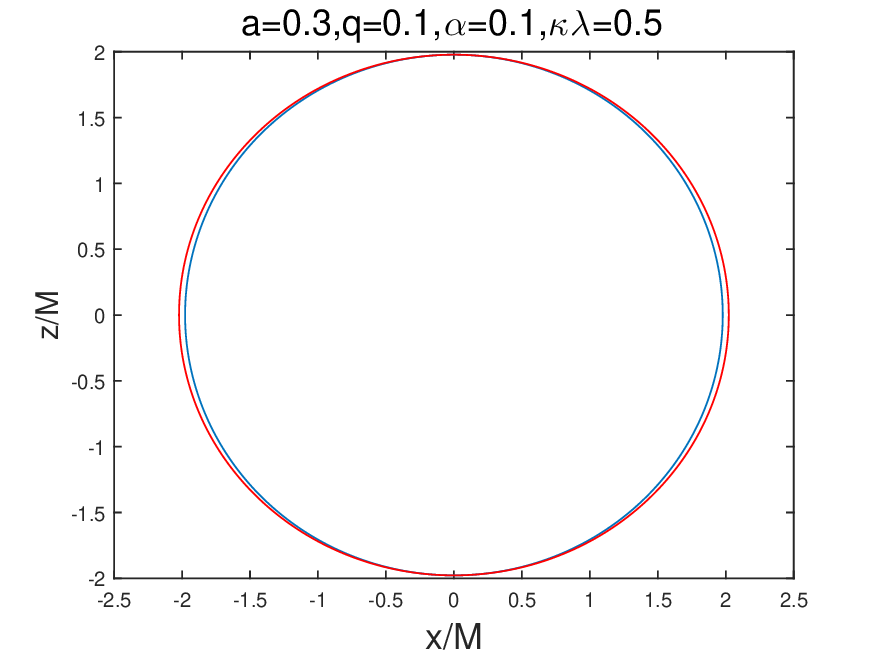}
  \includegraphics[scale=0.36]{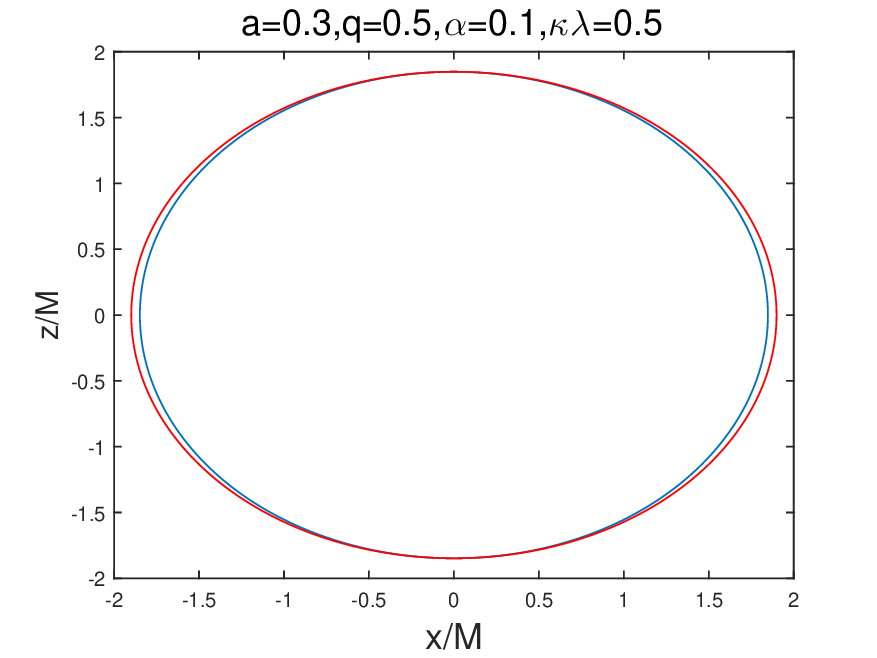}
  \includegraphics[scale=0.36]{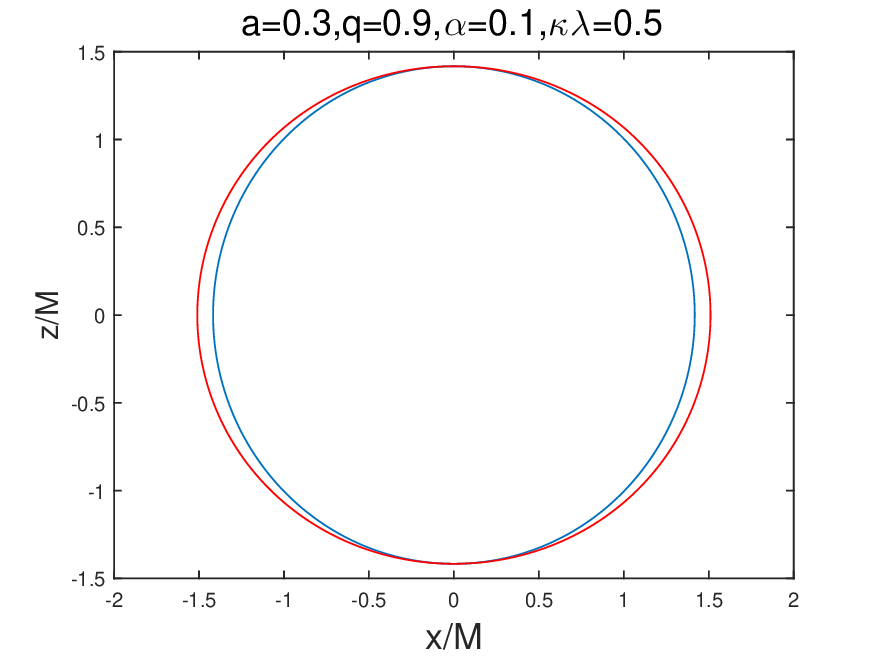}
  \includegraphics[scale=0.36]{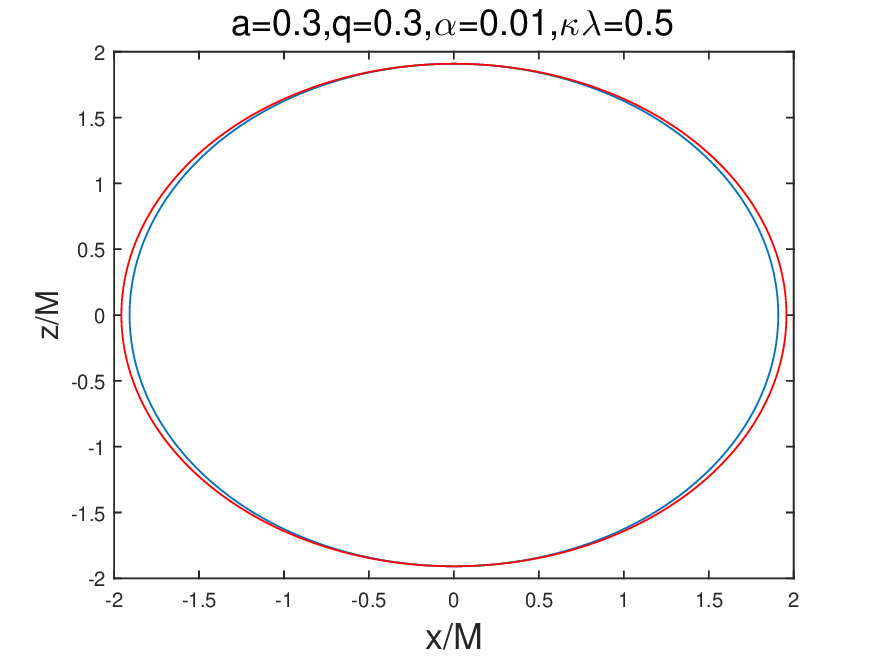}
  \includegraphics[scale=0.36]{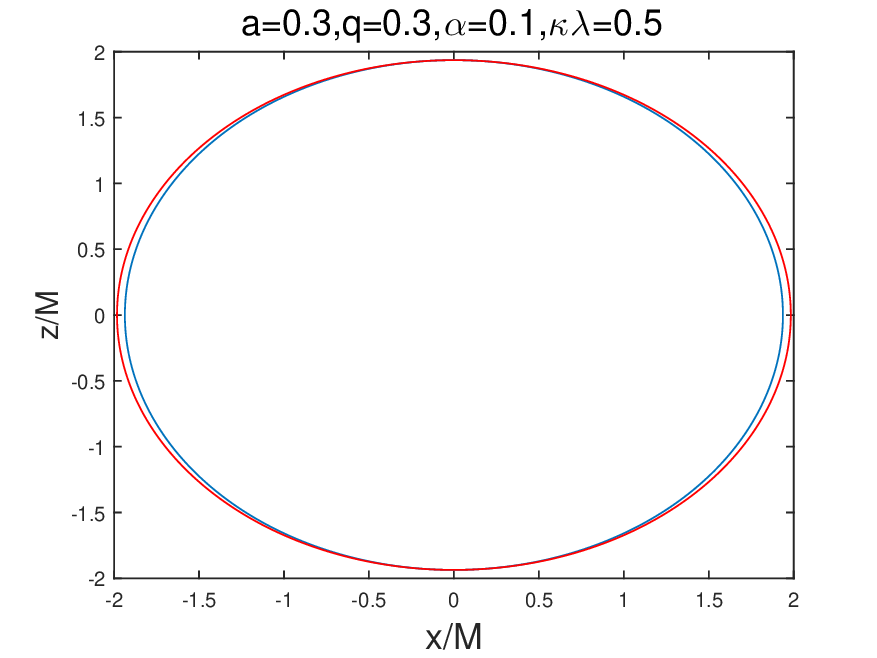}
  \includegraphics[scale=0.36]{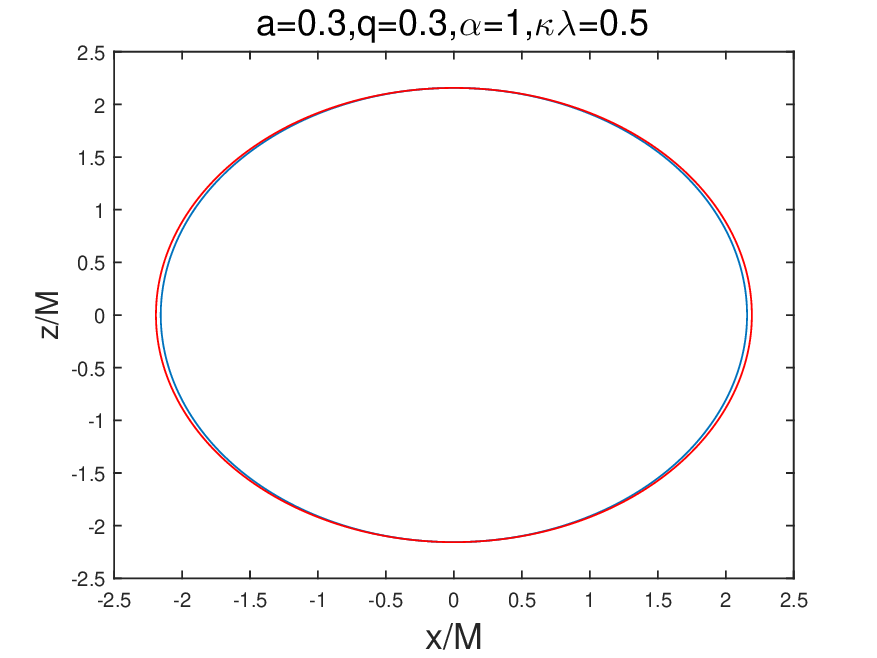}
  \includegraphics[scale=0.36]{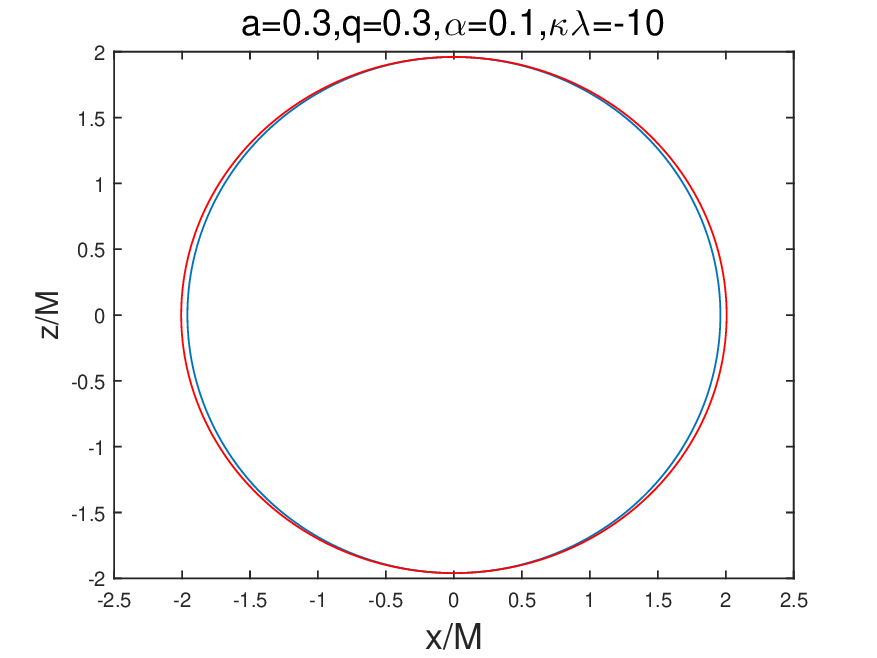}
  \includegraphics[scale=0.36]{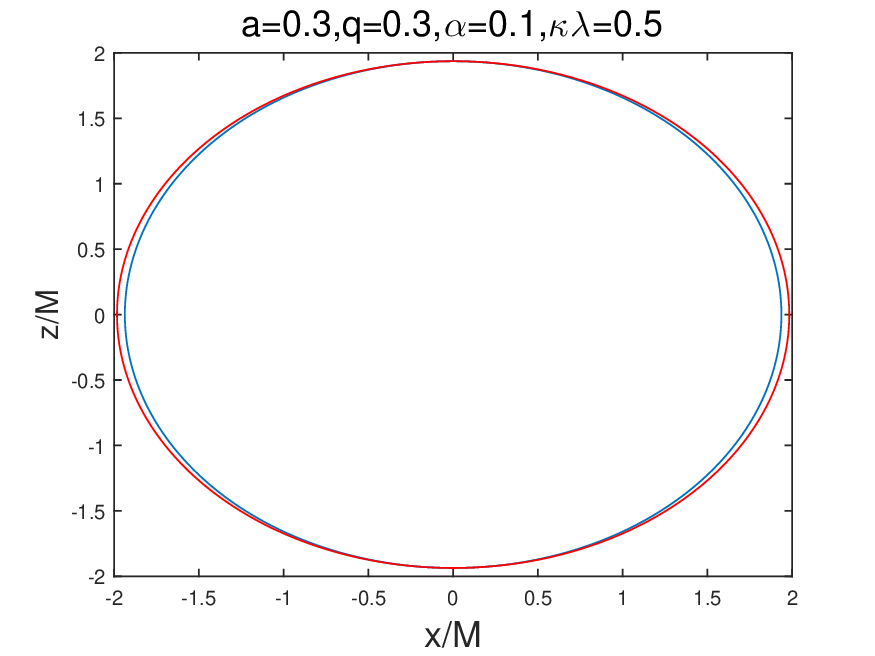}
  \includegraphics[scale=0.36]{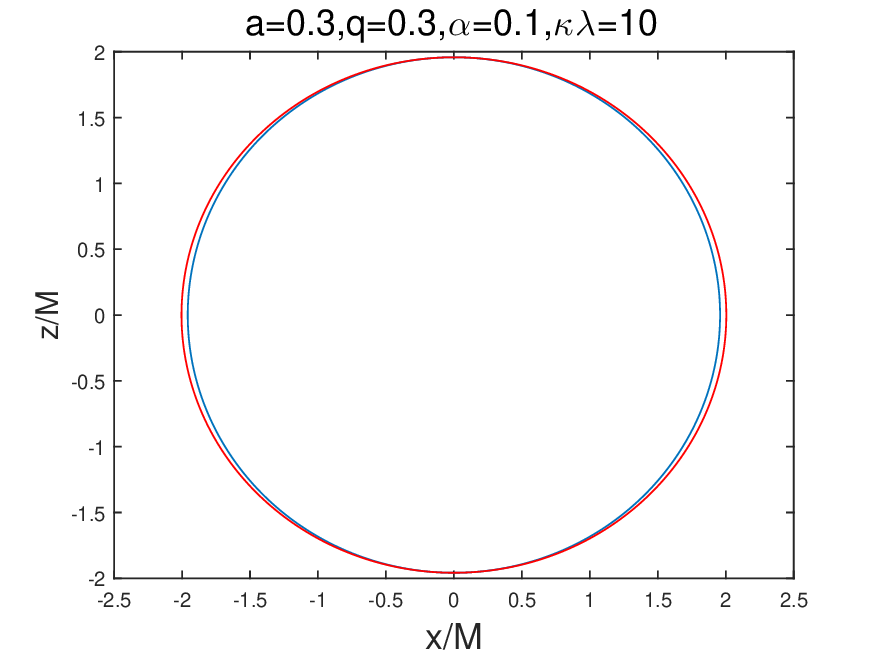}

   \caption{The ergosphere areas of the KN-AdS black hole surrounded by dark energy ($\omega=-2/3$) in the Rastall gravity for different $a, Q, \alpha, \kappa\lambda$.  The ergosphere region is between the event horizon (blue line) and the stationary limit surface (red line). The cosmological constant is set as $\Lambda\simeq 0$}
  \label{fig:1}
\end{figure}

\begin{figure}[htbp]
  \centering
  \includegraphics[scale=0.36]{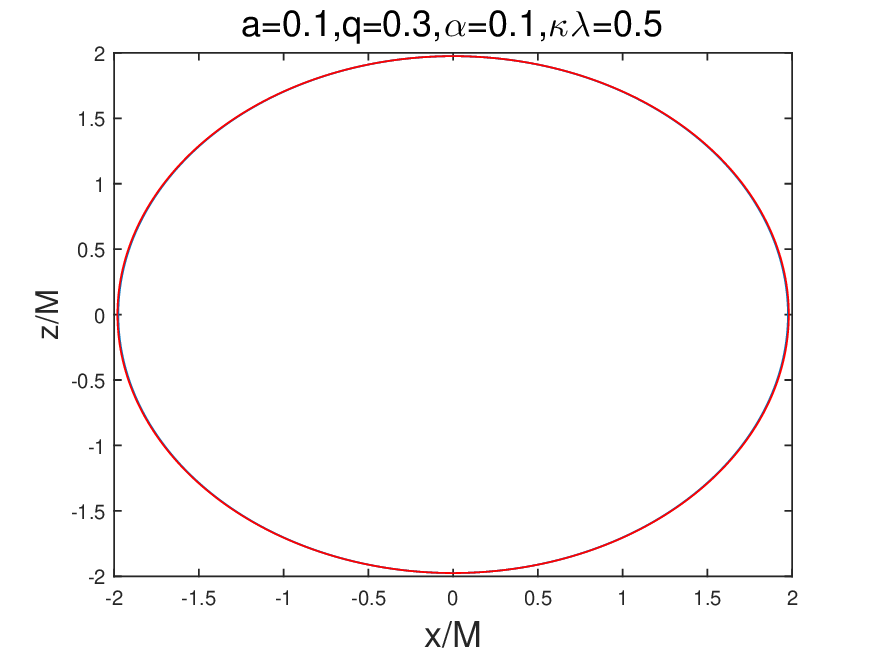}
  \includegraphics[scale=0.36]{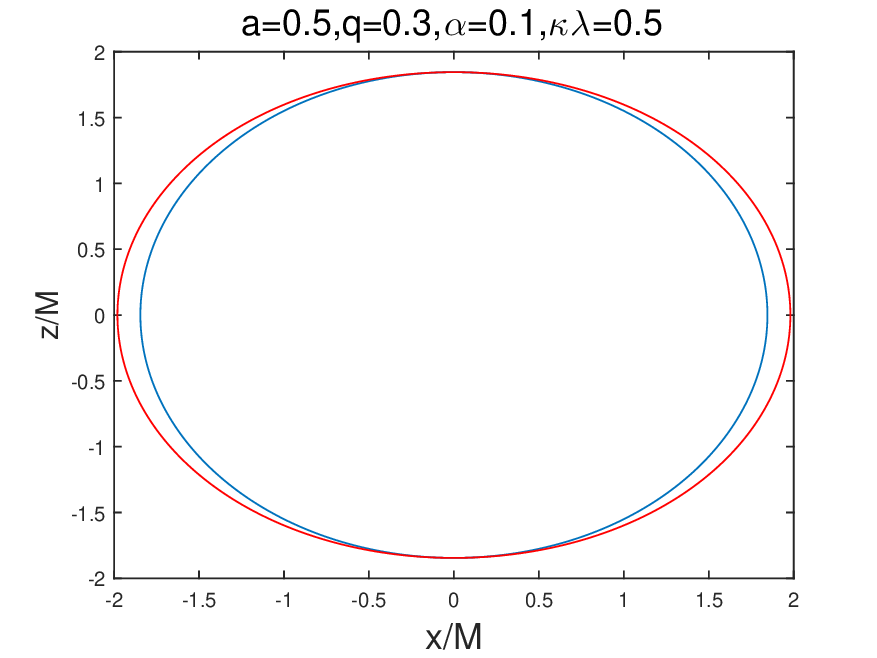}
  \includegraphics[scale=0.36]{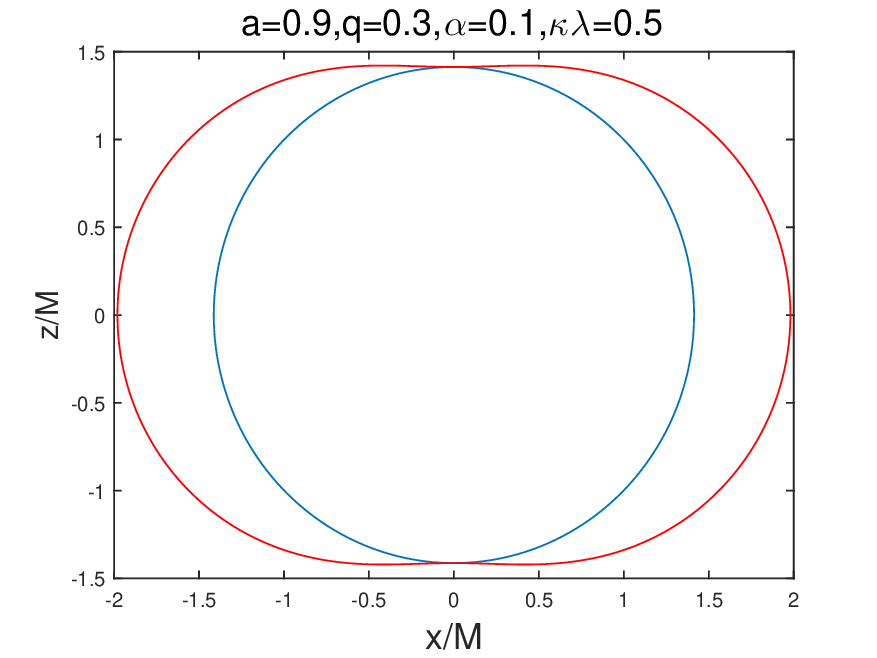}
  \includegraphics[scale=0.36]{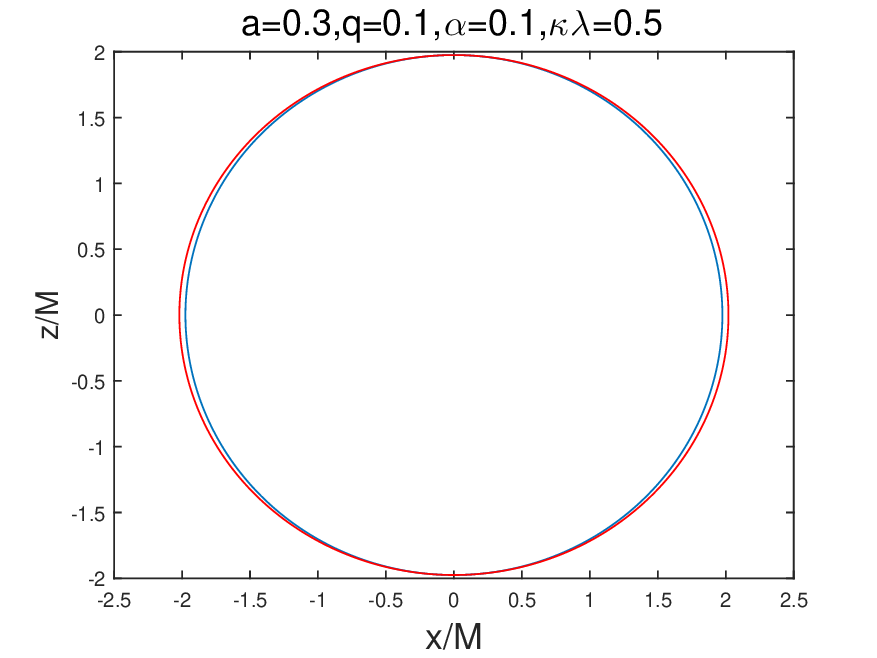}
  \includegraphics[scale=0.36]{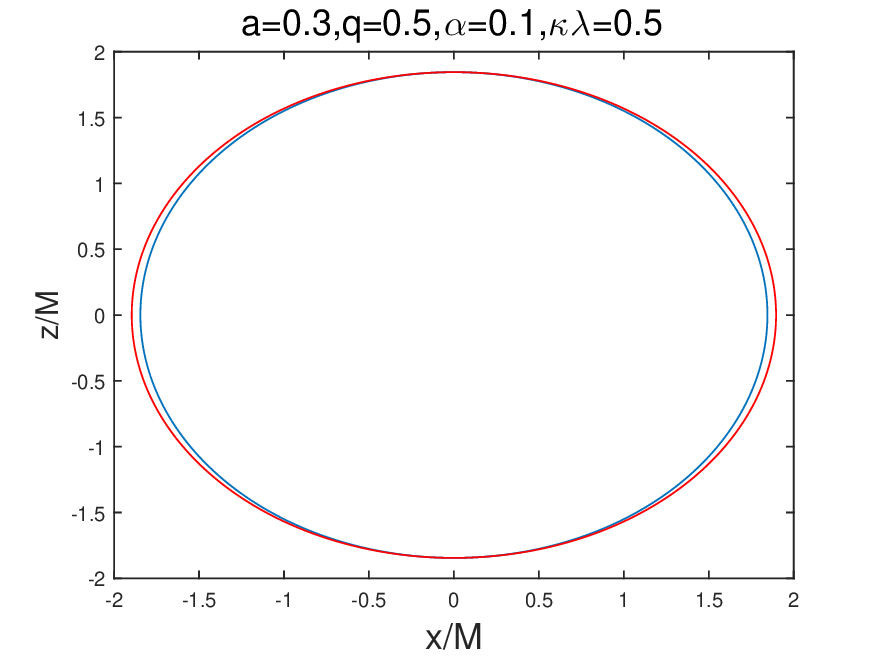}
  \includegraphics[scale=0.36]{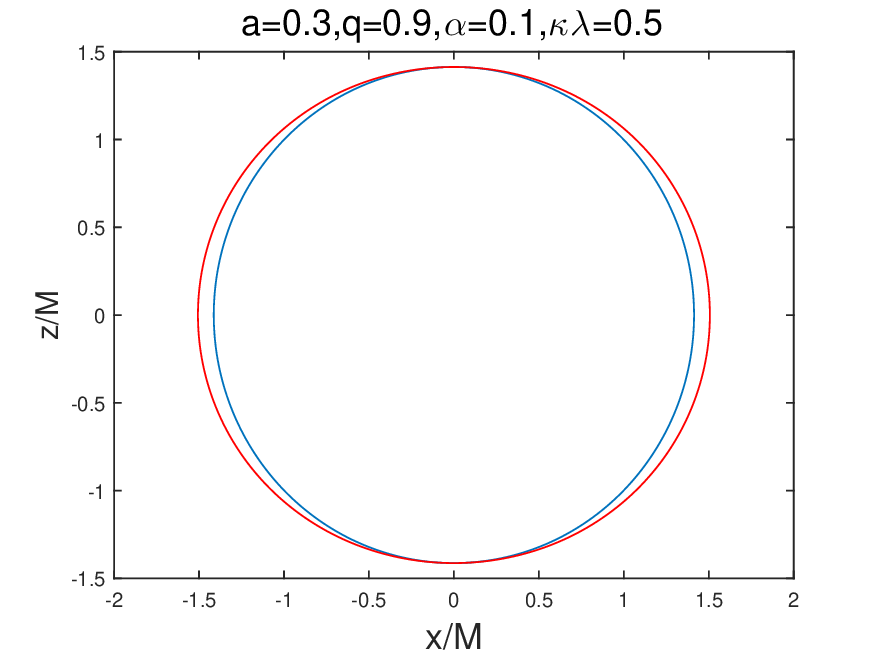}
  \includegraphics[scale=0.36]{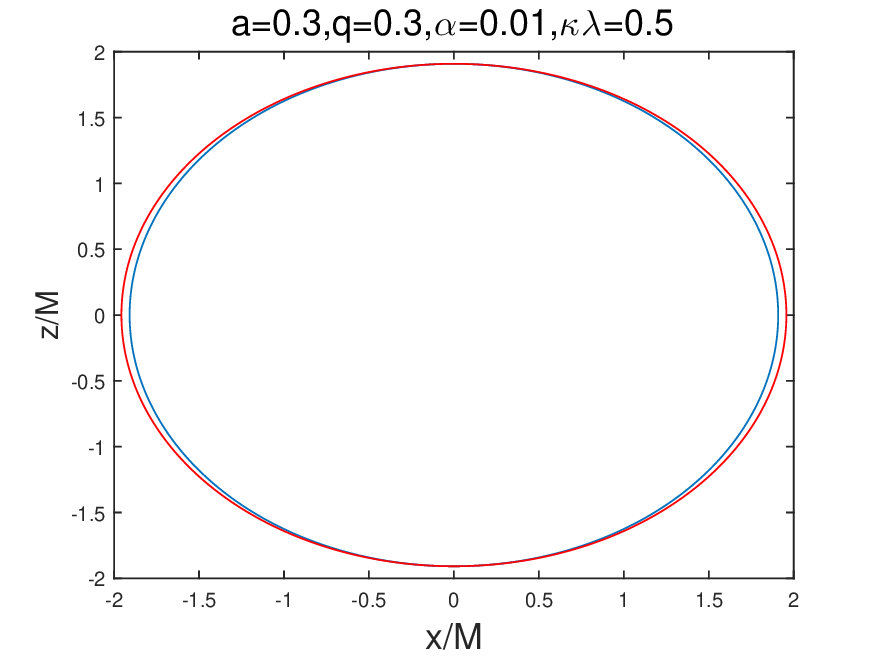}
  \includegraphics[scale=0.36]{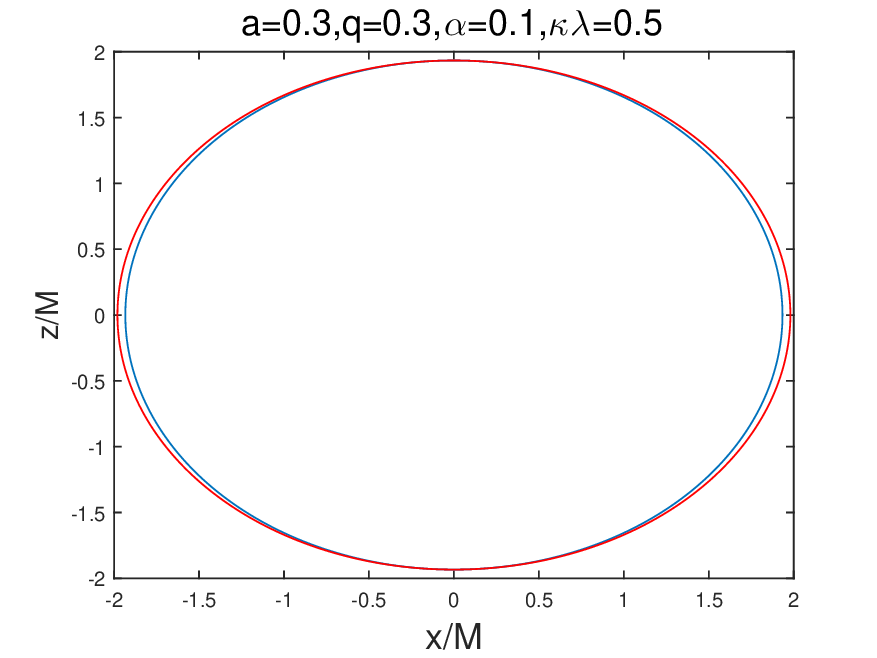}
  \includegraphics[scale=0.36]{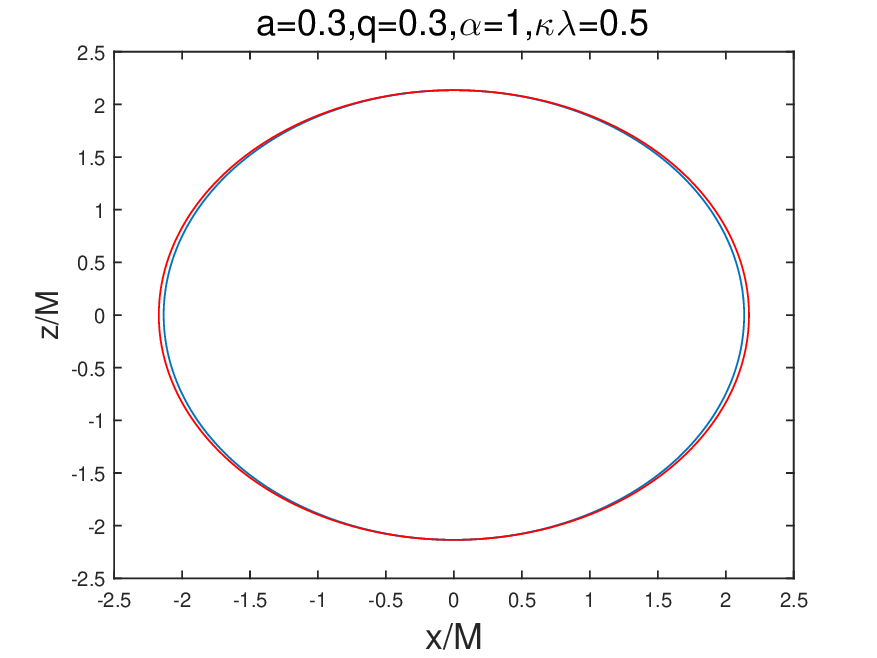}
  \includegraphics[scale=0.36]{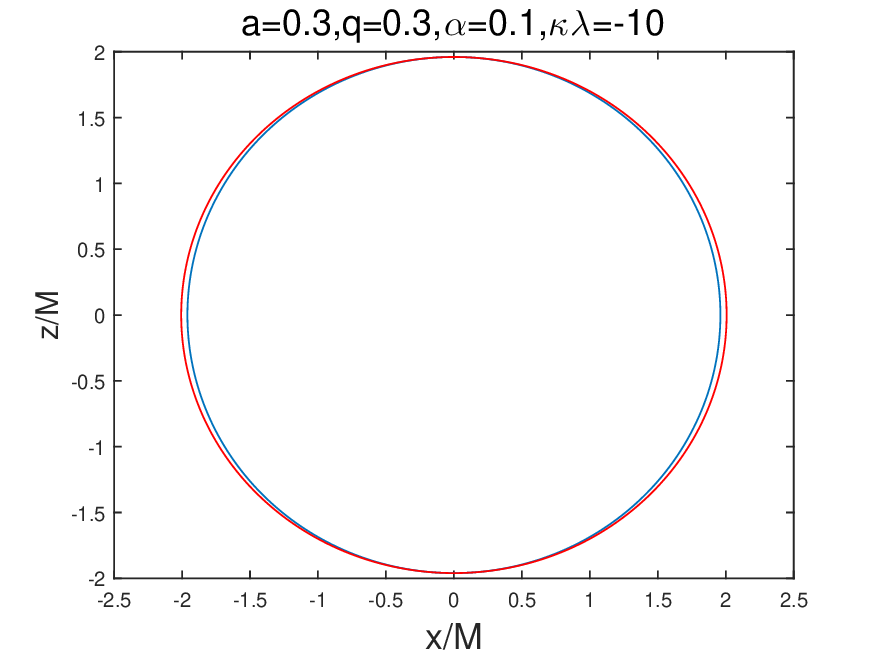}
  \includegraphics[scale=0.36]{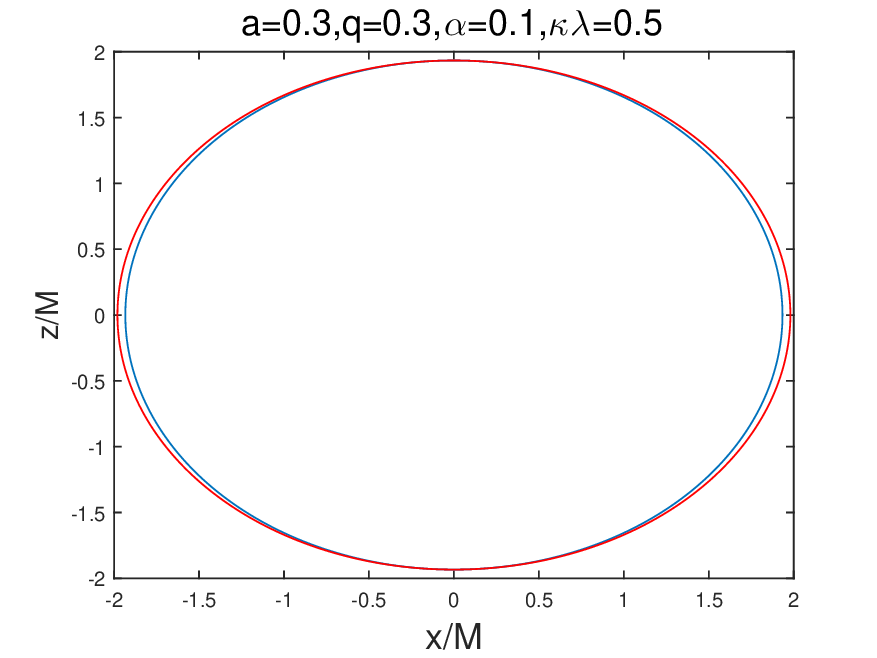}
  \includegraphics[scale=0.36]{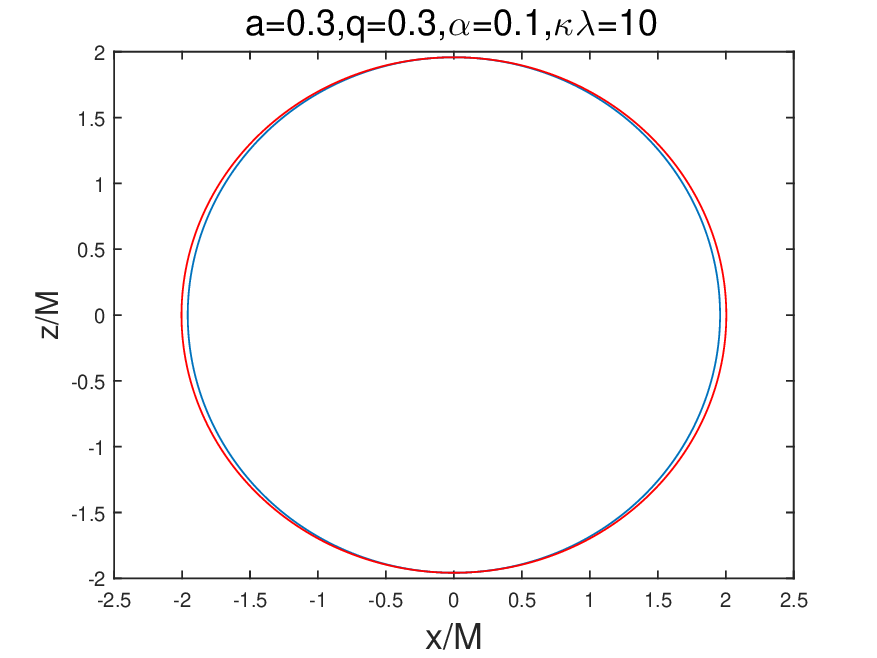}

   \caption{The ergosphere areas of the KN-AdS black hole surrounded by perfect fluid dark matter ($\omega=-1/3$) in the Rastall gravity for different $a, Q, \alpha, \kappa\lambda$. The ergosphere region is between the event horizon (blue line) and the stationary limit surface (red line). The cosmological constant is set as $\Lambda\simeq 0$}
  \label{fig:2}
\end{figure}

\subsection{Stationary limit surfaces}
The stationary limit surface of the KN-AdS black hole is important because the rotational energy of black hole relates to the ergosphere size. From the black hole metric (Eq. \ref{AdS5}), the ergosphere is defined by
\begin{equation}
g_{tt}=\dfrac{1}{\Sigma^{2}\Xi^{2}}(a^{2}sin^{2}\theta\Delta_{\theta}-\Delta_{r})=[\dfrac{\Lambda}{3}r^{2}(r^{2}+a^{2})-a^{2}cos^{2}\theta+$$$$
\dfrac{\Lambda}{3}a^{4}sin^{2}\theta cos^{2}\theta-r^{2}+2Mr-Q^{2}+\alpha r^{\dfrac{1-3\omega}{1-3\kappa\lambda(1+\omega)}}]\dfrac{1}{\Sigma^{2}\Xi^{2}}=0.
\label{SL1}
\end{equation}
Through calculation, Eq. (37) becomes
\begin{equation}
\dfrac{\Lambda}{3}r^{2}(r^{2}+a^{2})-a^{2}cos^{2}\theta+\dfrac{\Lambda}{3}a^{4}sin^{2}\theta cos^{2}\theta-r^{2}+2Mr-Q^{2}+\alpha r^{\dfrac{1-3\omega}{1-3\kappa\lambda(1+\omega)}}=0.
\label{SL2}
\end{equation}

This equation has two roots. There is an ergosphere between the event horizon and the static limit surface. The parameters $a, \omega, Q, \alpha, \Lambda, \theta$ and $\kappa\lambda$ could affect the shape of the ergosphere. In Fig. 1 and Fig. 2, we show the shapes of the ergosphere for dark energy ($\omega=-2/3$) and perfect fluid dark matter ($\omega=-1/3$), respectively. We find following properties of the ergosphere: (1) Its size decreases with the increasing $\alpha$, indicating that dark energy and perfect fluid dark matter reduce the rotational energy of the black hole; (2) Its size increases with the increasing $\kappa\lambda$ in the effective range. These results imply that the Mach principle or the re-arrangement of matter makes the rotational energy of the black hole larger.

\subsection{Singularity}
According to the Einstein GR, the stationary axisymmetric black hole space-time has the singularity. We consider whether or not the singularity changes with the presence of the dark energy and/or the perfect fluid dark matter. From the definition of the singularity, the Kretsmann scalar for all $\omega$ is given by
\begin{equation}
R=R^{\mu\nu\rho\sigma}R_{\mu\nu\rho\sigma}=\dfrac{H(r,\theta,a,\omega,\Lambda,Q^{2})}{\Sigma^{12}},
\label{S1}
\end{equation}
where $H$ is the polynomial function. The singularity occurs at $r=0$ and $\theta=\pi/2$ when $\Sigma^{2}=r^{2}+a^{2}cos^{2}\theta=0$. In the Boyer-Lindquist coordinates, the singularity is a ring with a radius of $a$ in the equatorial plane. For different $\omega$ and $\alpha$, there are different polynomial functions, but $\Sigma^{2}$ is the same as in the usual Kerr black hole, so the perfect fluid dark matter and dark energy do not change the singularity of the black hole.

\section{ROTATION VELOCITIES IN THE EQUATORIAL PLANE AND ROTATION CURVE DIVERSITY}
Basing on the KN-AdS solution in perfect fluid matter in the Rastall gravity (Section 3), we now discuss the rotation curves in the equatorial plane ($\theta=\dfrac{\pi}{2}$). We first derive the rotation velocity expression using ordinary method, then we calculate the rotation velocities with the presence of the dark energy and perfect fluid dark matter. Especially for the case of perfect fluid dark matter, we can explain the diversity of rotation curve if considering both the perfect fluid dark matter and a baryon disk which is expressed as $\rho_{b}=\Sigma_{0}exp[-r/r_{d}]$, where $\Sigma_{0}$ is the central surface density and $r_{d}$ is the scale radius.

In the four-dimensional space-time, the four-velocity satisfies the normalized condition for zero angular momentum observer (ZAMO) as $g_{\mu\nu}u^{\mu}u^{\nu}=-1$. In the axisymmetric space-time, the space-time symmetry leads to two conserved quantities $L$ and $E$. In our KN-AdS black hole space-time, we write the normalized condition from the expressions of $u^{\mu}$ and $u^{\nu}$ as
\begin{equation}
g_{tt}(\dfrac{dt}{d\tau})^{2}+2g_{t\phi}\dfrac{dt}{d\tau}\dfrac{d\phi}{d\tau}+g_{\phi\phi}(\dfrac{d\phi}{d\tau})^{2}+g_{rr}(\dfrac{dr}{d\tau})^{2}=-1.
\label{RC1}
\end{equation}
We then obtain 
\begin{equation}
-E\dfrac{dt}{d\tau}+L\dfrac{d\phi}{d\tau}+g_{rr}(\dfrac{dr}{d\tau})^{2}=-1.
\label{RC2}
\end{equation}
Combining the equations of (\ref{RC1}) and (\ref{RC2}), we get
\begin{equation}
(\dfrac{dr}{d\tau})^{2}=-\dfrac{1}{g_{rr}}+\dfrac{g_{\phi\phi}E^{2}+2g_{t\phi}EL+g_{tt}L^{2}}{(g^{2}_{t\phi}-g_{tt}g_{\phi\phi})g_{rr}}=E^{2}-V^{2}.
\label{RC3}
\end{equation}
In the black hole space-time, the general conditions for a stable circular orbit are $\dfrac{dr}{d\tau}=0$ and $\dfrac{dV^{2}}{dr}=0$, we then get the expressions of $L$ and $E$ as
\begin{equation}
E=\pm\dfrac{g_{tt}+g_{t\phi}\Omega_{\phi}}{\sqrt{-g_{tt}-2g_{t\phi}\Omega_{\phi}-g_{\phi\phi}\Omega^{2}_{\phi}}},$$$$
L=\pm\dfrac{g_{t\phi}+g_{\phi\phi}\Omega_{\phi}}{\sqrt{-g_{tt}-2g_{t\phi}\Omega_{\phi}-g_{\phi\phi}\Omega^{2}_{\phi}}},$$$$
\Omega_{\phi}=\dfrac{-g_{t\phi,r}+\sqrt{(g_{t\phi,r})^{2}-g_{tt,r}g_{\phi\phi,r}}}{g_{\phi\phi,r}}.
\label{RC4}
\end{equation}
From the definition of rotation velocity, we obtain
\begin{equation}
v=\dfrac{L}{\sqrt{g_{\phi\phi}}}=\dfrac{1}{\sqrt{g_{\phi\phi}}}\dfrac{g_{t\phi}+g_{\phi\phi}\Omega_{\phi}}{\sqrt{-g_{tt}-2g_{t\phi}\Omega_{\phi}-g_{\phi\phi}\Omega^{2}_{\phi}}}
\label{RC5}
\end{equation}

\begin{figure}[htbp]
  \centering
  \includegraphics[scale=0.5]{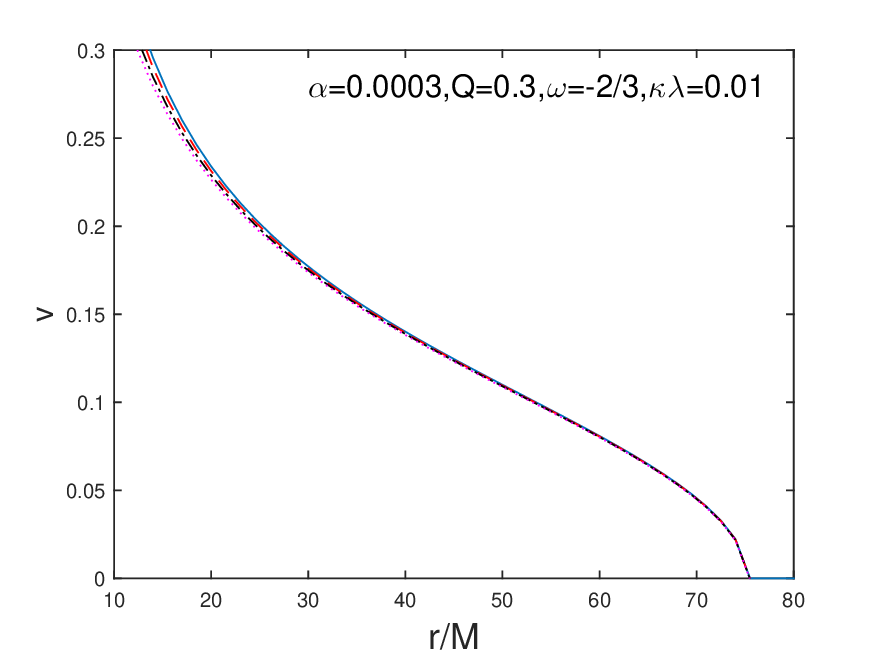}
  \includegraphics[scale=0.5]{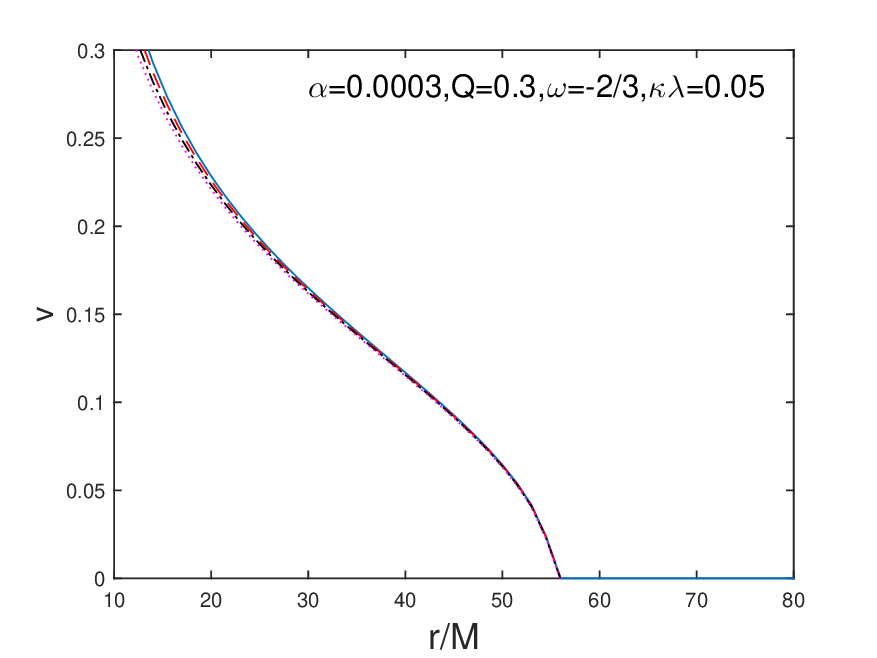}
  \includegraphics[scale=0.5]{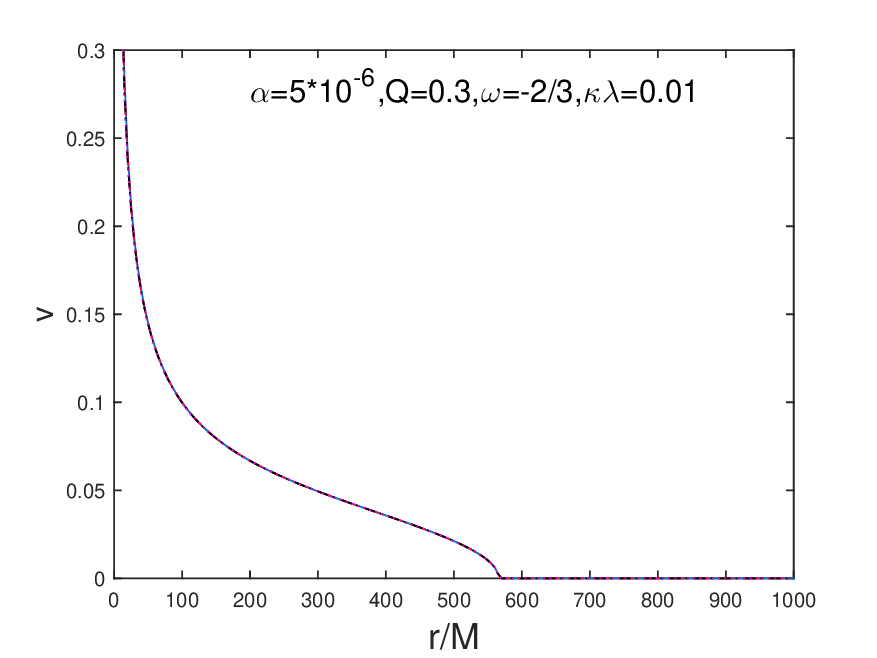}
  \includegraphics[scale=0.5]{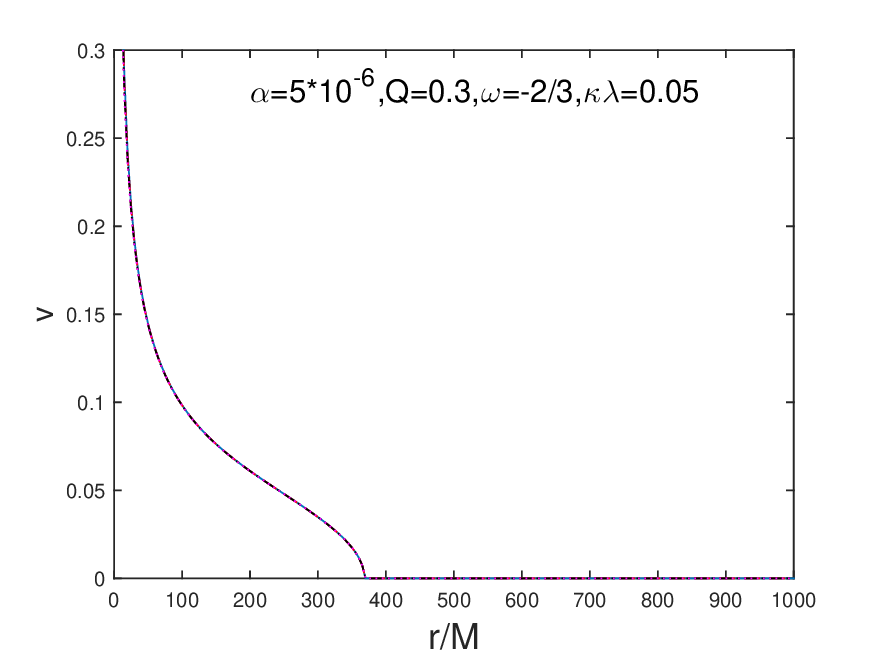}
   \caption{Curves of the rotation velocity for the KN-AdS black hole surrounded by the dark energy in the Rastall gravity.}
  \label{fig:3}
\end{figure}

\begin{figure}[htbp]
  \centering
  \includegraphics[scale=0.36]{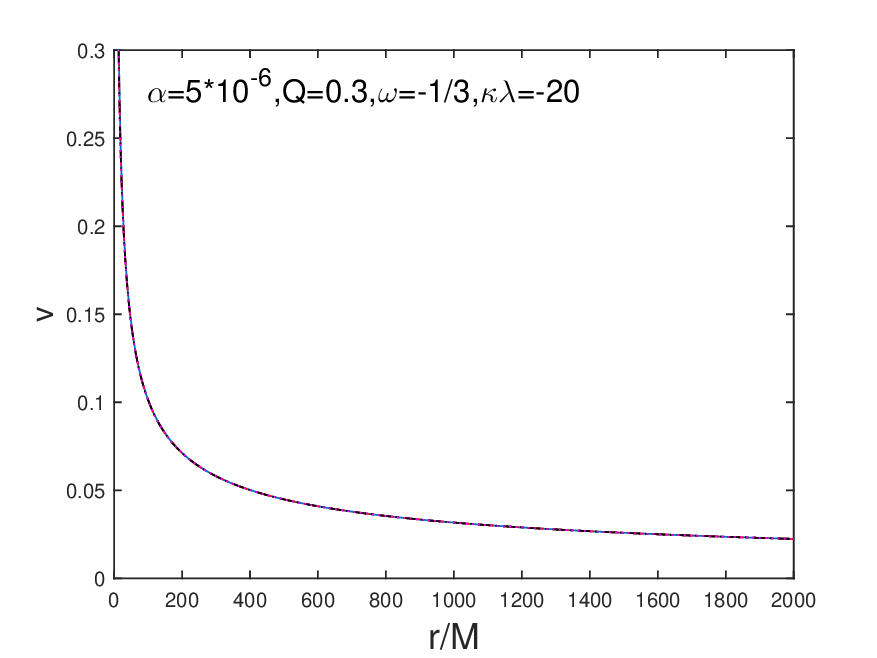}
  \includegraphics[scale=0.36]{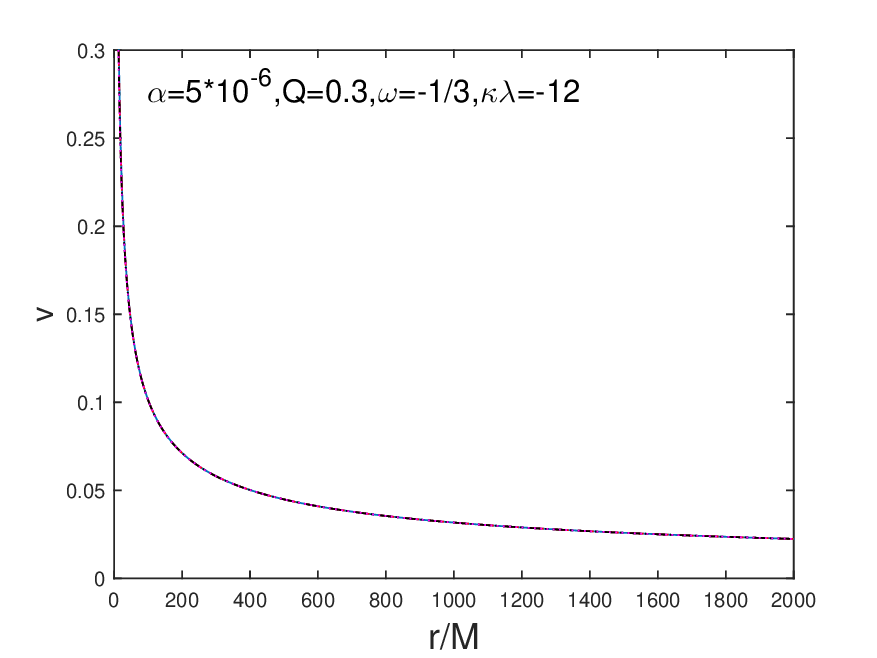}
  \includegraphics[scale=0.36]{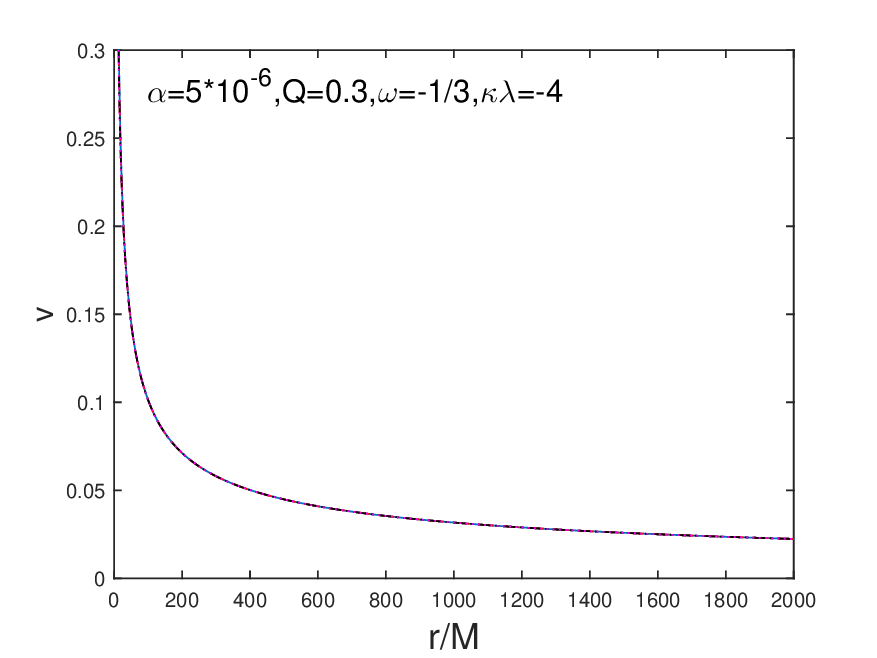}
  \includegraphics[scale=0.36]{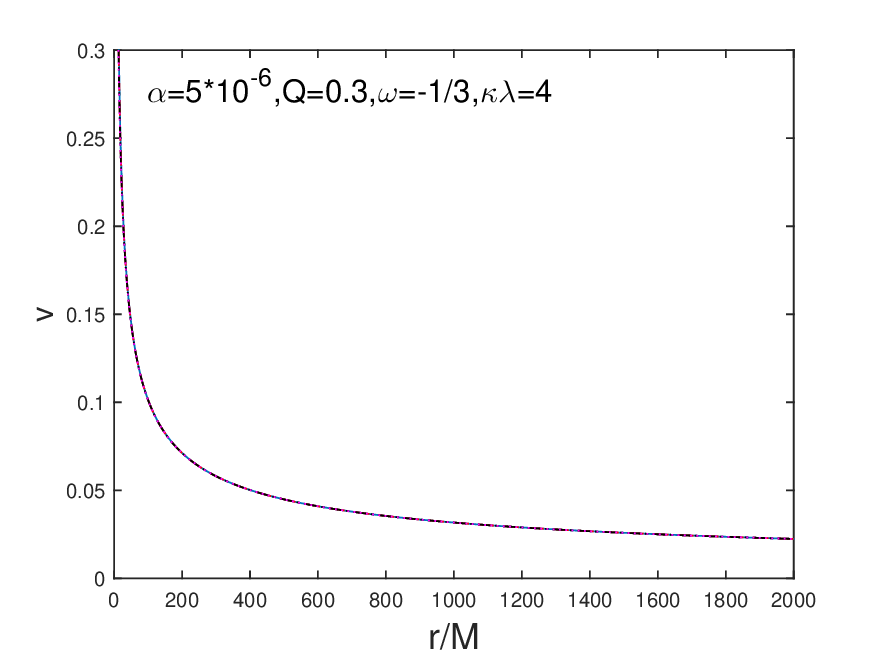}
  \includegraphics[scale=0.36]{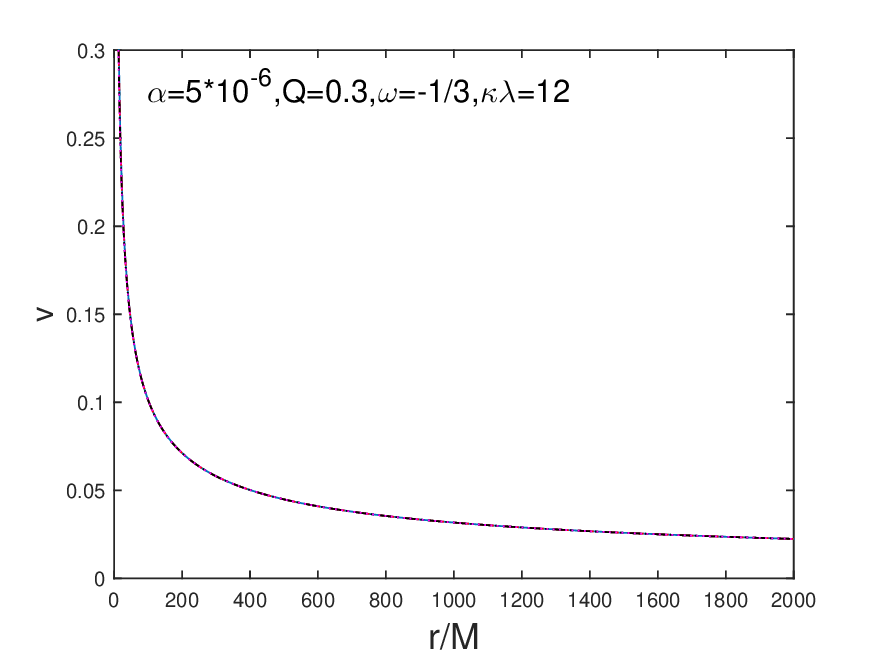}
  \includegraphics[scale=0.36]{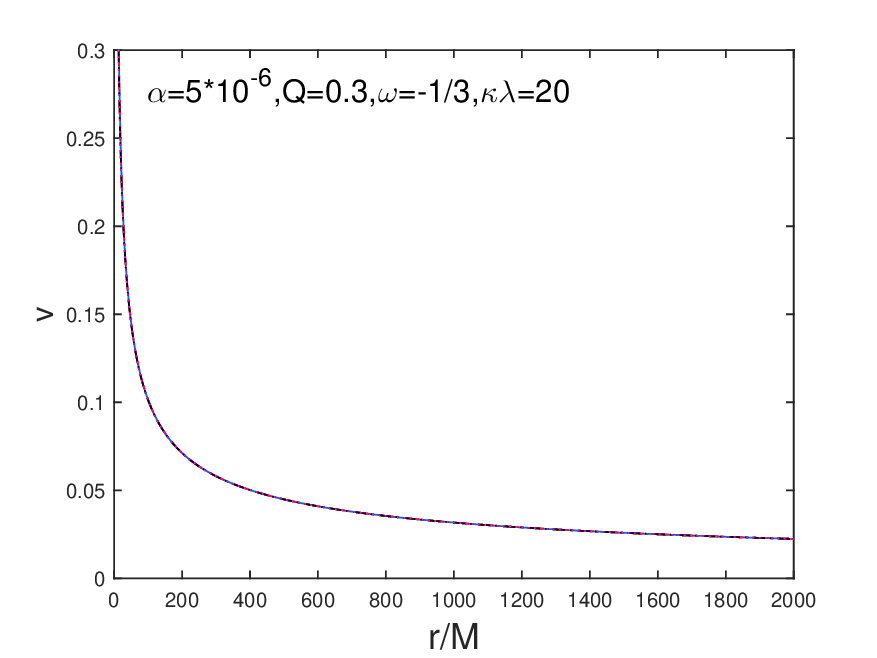}
   \caption{Curves of the rotation velocity for the KN-AdS black hole surrounded by the perfect fluid dark matter in the Rastall gravity.}
  \label{fig:4}
\end{figure}

Fig. 3 and Fig. 4 show the curves of the rotation velocity for the KN-AdS black hole surrounded by the dark energy and perfect fluid dark matter, respectively, in the Rastall gravity. We find that: (1) In the case of dark energy, the rotation velocity always decreases with the increasing distance from the black hole and becomes flatter with the decreasing $\kappa\lambda$. (2) In the case of perfect fluid dark matter, the influence of $\kappa\lambda$ on the rotation velocity is minor at small distances and significant at larger distances. Note that the effect at larger distances is not shown in the figures, but can be inferred from the explicit expression of the rotation velocity (Eq. 46).

As an example of the application of the Rastall gravity, we try to explain the rotation curve diversity of Low Surface Brightness (LSB) galaxies. In the Rastall gravity, the parameter $\kappa\lambda$ represents the re-arrangement of the perfect fluid matter around the black hole. From Section 3, for the perfect fluid dark matter halo ($\omega=-1/3$), we can calculate the energy density $\rho$ through the Einstein equation. Because the motion velocity of the dark matter particle is much smaller than the speed of light, the energy density of the perfect fluid dark matter then approximates to the mass density. On the other hand, the rotation curve is usually observed at large distances from the black hole, so that the black hole spin $a$ then can be approximated to zero. From \cite{2017PhRvL.119k1102K}, the baryon matter can be considered as an index disk, i.e., $\rho_{b}=\Sigma_{0}exp[-r/r_{d}]\delta(z),$ in galaxies. Using the perfect fluid dark matter halo and the baryon disk, we can derive the rotation curve expression. In the equatorial plane, the general Mass function is $M(r)=4\pi\int^{r}_{0} r^{2}\rho_{DM} dr+2\pi\int^{r}_{0} r\rho_{b} dr  $. We then obtain the Mass function in our case as
\begin{equation}
M(r) =\frac{\alpha}{2} \frac{1-4\kappa\lambda}{1-2\kappa\lambda}r^{\frac{2\kappa\lambda+1}{1-2\kappa\lambda}}-2\pi\Sigma_{0}r_{d}exp[\dfrac{-r}{r_{d}}](r_{d}+r).
\label{RC6}
\end{equation}
Thus the rotation velocity $v$ in the equatorial plane is given by
\begin{equation}
v=\sqrt{\dfrac{GM(r)}{r}}=\sqrt{\frac{G\alpha}{2} \frac{1-4\kappa\lambda}{1-2\kappa\lambda}r^{\frac{4\kappa\lambda}{1-2\kappa\lambda}}-2\pi G\Sigma_{0}r_{d}exp[\dfrac{-r}{r_{d}}](\dfrac{r_{d}}{r}+1)}.
\label{RC7}
\end{equation}
The rotation velocity is determined by $\alpha, \Sigma_{0}, r_{d}$ and $\kappa\lambda$. Varying $\kappa\lambda$ would result in different expressions of $v$, which may explain the rotation velocity diversity in LSB galaxies. We stress two points on the Rastall gravity. If the Rastall gravity is equivalent to the Einstein GR, the parameter $\kappa\lambda$ only describes the re-arrangement of the perfect fluid dark matter. If the Rastall gravity is not equivalent to the Einstein GR, the meaning of $\kappa\lambda$ is the same, but the physical origin of $\kappa\lambda$ is related to the Mach principle.

\section{SUMMARY}
We have obtained the KN-AdS black hole solutions surrounded by perfect fluid matter in the Rastall gravity using the Newman-Janis method and complex calculations. In this work, we focus on the cases of the dark energy ($\omega=-2/3$) and the perfect fluid dark matter ($\omega=-1/3$). We discuss different black hole properties in the Rastall gravity with the presence of the perfect fluid matter. We find that if the condition $(3\kappa\lambda(1+\omega)-3\omega)(1-4\kappa\lambda)\geq 0$ is satisfied, the WEC and SEC can always be hold everywhere in the KN-AdS black hole. By analyzing the horizon equation, we can constrain the values of the parameter $\alpha$, which is determined by the equation of state $\omega$ and the Rastall parameter $\kappa\lambda$. In the case of the dark energy, the maximum value of $\alpha$ decreases with the increasing $\kappa\lambda$ when $0<\kappa\lambda$ and $\kappa\lambda\neq 1$, and increases with the increasing $\mid\kappa\lambda\mid$ when $-2<\kappa\lambda<0$. In the case of the perfect fluid dark matter, the maximum value of $\alpha$ decreases with the increasing $\kappa\lambda$ when $0<\kappa\lambda<1$ and $\kappa\lambda\neq 1/2$, and increases with the increasing $\mid\kappa\lambda\mid$ when $\kappa\lambda<0$. The perfect fluid dark matter and dark energy reduce the ergosphere size, but they don't change the singularity of the black hole.

We also derive the rotation velocity for the KN-AdS black hole surrounded by the perfect fluid matter in the Rastall gravity. In the case of dark energy, the rotation velocity decreases with the increasing distance from the black hole and becomes flatter with the decreasing $\kappa\lambda$. In the case of perfect fluid dark matter, the effect at small distances is minor and become significant at galactic scale. Considering both the perfect fluid dark matter halo and the baryon disk, we have the possibility of explaining the rotation curve diversity in LSB galaxies.

\acknowledgments
We acknowledge the anonymous referee for a constructive report that has significantly improved this paper. We acknowledge the financial support from the National Natural Science Foundation of China through grants 11503078, 11573060 and 11661161010.

\end{document}